\documentclass[twocolumn,aps,superscriptaddress,showpacs]{revtex4}

\usepackage{float}
\usepackage{amsmath}
\usepackage{amssymb}
\usepackage{esint}
\usepackage{graphicx}
\usepackage{epsfig}
\usepackage{dcolumn}
\usepackage{color}
\usepackage[english]{babel}
\usepackage{mathrsfs}
\usepackage{bm}
\usepackage{lscape}

\begin{document}

\title{ Topological aspects in the photonic crystal analog \\ of single-particle transport in quantum Hall systems }

\author{Luca Esposito} 
\email{esposito.luca@mail.com}
\affiliation{Department of Physics, University of Pavia, via A. Bassi 6, I-27100 Pavia, Italy}

\author{Dario Gerace}
\affiliation{Department of Physics, University of Pavia, via A. Bassi 6, I-27100 Pavia, Italy}

\begin{abstract}
We present a perturbative approach to derive the semiclassical equations of motion for the two-dimensional
electron dynamics under the simultaneous presence of static electric and magnetic fields, where the quantized 
Hall conductance is known to be directly related to the topological properties of translationally invariant magnetic 
Bloch bands. In close analogy to this approach, we develop a perturbative theory of two-dimensional photonic 
transport in gyrotropic photonic crystals to mimic the physics of quantum Hall systems. We show that a suitable 
permittivity grading of a gyrotropic photonic crystal is able to simulate the simultaneous presence of analog electric
and magnetic  field forces for photons, and we rigorously derive the topology-related term in the
equation for the electromagnetic energy velocity that is formally equivalent to the electronic case.
A possible experimental configuration is proposed to observe a bulk photonic analog to the quantum Hall physics 
in graded gyromagnetic photonic crystals.

\end{abstract}


\maketitle

\section{Introduction}

Since its first phenomenological observation more than thirty years ago \cite{key-1,tsui}, 
the physics of the quantum Hall effects has spurred a wealth of groundbreaking 
theoretical achievements, which have eventually clarified the generality of the topological aspects at the heart of this 
fascinating problem \cite{phase_book,chang_review}.  It is now understood that the dynamical properties of 
the two-dimensional (2D) electron motion under the simultaneous presence of electric and magnetic
fields are determined by a topological invariant of the Bloch bands, an integer known as the Chern 
number \cite{key-16}, which is different from zero only after time-reversal symmetry (TRS) is broken by the 
external magnetic field perpendicular to the plane of motion. 
As a consequence, the semiclassical equations of motion for the electron group velocity depend 
on a topological term related to the non-vanishing Berry curvature \cite{key-17,key-18,key-19}.
The relevance of such topological theories is twofold. 
On one hand, the generality of geometrical properties has been extensively used to explain a number of 
physical phenomena in condensed matter, from the anomalous Hall effect \cite{key-2} 
to the existence of topological superconductors and insulators \cite{key-4}. 
On the other, since the topological invariant is a global property of the energy eigenstates of the system, 
it is intrinsically robust against system perturbations, such as lattice distortions and disorder. 
As a typical example, in a quantum Hall system the transverse conductance is a multiple of the Chern 
invariant of the gauge bundle \cite{key-15,key-16} , for which its value is extremely 
stable against structural characteristics of the system, and it is measured with accuracies of one part on 
hundred million \cite{key-8}.  As a further consequence, topologically non-trivial systems possess 
chiral ballistic edge states at the border of a finite sample \cite{key-9,key-10}. 
Such states, induced by the spatial boundary between systems with distinct topological phases, allow 
uni-directional and nonreciprocal electronic transport 
\cite{foot1}, and they are intrinsically immune to back-scattering. 

The analogies between photonic band dispersion in artificially periodic electromagnetic
systems, known as photonic crystals \cite{key-14}, and the electron band theory in crystalline solids have 
recently motivated the idea that TRS breaking allows non-trivial
topological properties of the photonic modes in such systems \cite{key-13,key-5}. 
Typically, Faraday-active elements arranged in a periodic lattice produce the required breaking of symmetry, 
necessary to induce a non-vanishing Chern number for photonic bands \cite{key-11}. 
Following these early proposals, propagation of back-scattering immune photonic edge states has been 
observed at the interface between a magneto-optical photonic crystal and
a topologically trivial photonic medium \cite{key-12}.
Clearly, these features could be very important for future applications in integrated photonic circuits, 
because of the possibility to exploit uni-directional channels of electromagnetic energy transport that are 
intrinsically insensitive to disorder in the sample, just like electronic transport in quantum Hall systems. 
More recently, several theoretical works have elaborated on the topological nature of one-way photonic
edge modes in specific gyroelectric \cite{yannopapas,ochiai2010,fang2011prb,yannopapas2012} photonic crystals,
TRS breaking in microwave circuits \cite{koch2010,Hafezi2011}, or the generation of artificial gauge fields
for photons in coupled cavity arrays \cite{carusot2011pra,fang2012prl,fang2012nphot}. 
The photonic analog of topological insulators have also been recently proposed \cite{khanikaevNmat2012} and
observed \cite{rechtsman2013}, along the same lines of previous works \cite{key-11,key-12}.
However, the theoretical problem of recovering the effective photon dynamics in TRS broken
photonic systems, in full analogy to the electron transport theory, has been not fully explored in the literature,
to our knowledge. A few early attempts to derive a topological-based photon dynamical theory were mostly focussed
on systems without TRS breaking \cite{onoda2004,onoda2006}, i.e. with a strict
analogy with the classical Hall transport properties. A rigorous derivation of the topological terms in
the semiclassical equations of motion for photonic transport starting from a in direct analogy between
Bloch-Floquet photonic modes and the magnetic Bloch electron states is still lacking.
 
Here we go beyond previous works in analyzing the analogies between electronic 
and photonic formalisms for TRS broken 2D crystals. 
To this end, we will first present a perturbative approach to obtain the equations 
of motion for the electron transport in quantum Hall systems, re-deriving the well known result 
that the semiclassical electron dynamics is described by \cite{key-17,key-18,key-19}
\begin{alignat}{1}
\mathbf{v}_{n\mathbf{k}} 
& =\frac{1}{\hslash}\nabla_{\mathbf{k}}E_{n\mathbf{k}}-\dot{\mathbf{k}}\times\boldsymbol{\Omega}_{n\mathbf{k}}\label{eq:2}\\
\dot{\mathbf{k}} 
& =-\frac{e\mathcal{E}}{\hslash}  \, , \label{eq:3}
\end{alignat}
where $\mathbf{k}$ is the wave vector, $n$ is the band index, $\mathbf{v}_{n}(\mathbf{k})$
is the group velocity associated with the magnetic Bloch band energy $E_{n}(\mathbf{k})$,
$\mathcal{E}$ the applied electric field, and 
$\boldsymbol{\Omega}_{n}(\mathbf{k})$
the Berry curvature of the gauge bundle constructed on the Brillouin zone.
Essentially, TRS breaking results in a topological correction, given by the Berry curvature,
to the standard equations of motion for the electron in the periodic potential of crystalline
solids (see, e.g., \cite{key-20} for a textbook-like formulation).
We will then apply the same formalism to Maxwell equations in periodic meta-materials with gyrotropic components 
and weak grading along one direction, rigorously obtaining the equation for the electromagnetic mode velocity 
containing an analogous topological correction, as already conjectured in \cite{key-5,key-13,key-11}.
As a final remark, we point out that in the present work we are mainly concerned with the link between linear 
photonic crystal theory and the topological aspects of single-electron transport in quantum Hall systems, 
while we are not dealing with the interesting problem of mimicking manybody quantum states, such as 
the ones leading to the fractional quantum Hall phenomenology \cite{tsui}, with in strongly nonlinear photonic 
systems \cite{carusot2012prl}.

The paper is organized as follows. In the first part, Sec.~II, we present
a perturbative approach to derive the known results of a
topological term in the single-electron semiclassical equations of motion in 
quantum Hall systems.
In the second part, Sec.~III, we explicitly treat photonic crystals on an analog footing, 
by applying the same perturbative concepts from Sec.~II to Maxwell equations.
We will then show that a combination of gyrotropic materials and
weak grading of the photonic crystal permittivity along the propagation 
direction are able to closely mimic the semiclassical single-electron dynamics
in quantum Hall systems also from a topological perspective.
Finally, in Sec.~IV we give some conclusive remarks, by proposing a possible experimental
setting where these geometrical aspects could be probed through photon transmission.

% FIG. 1
\begin{figure}[t]
\begin{center}
\vspace{-1.3cm}
\includegraphics[width=0.55\textwidth]{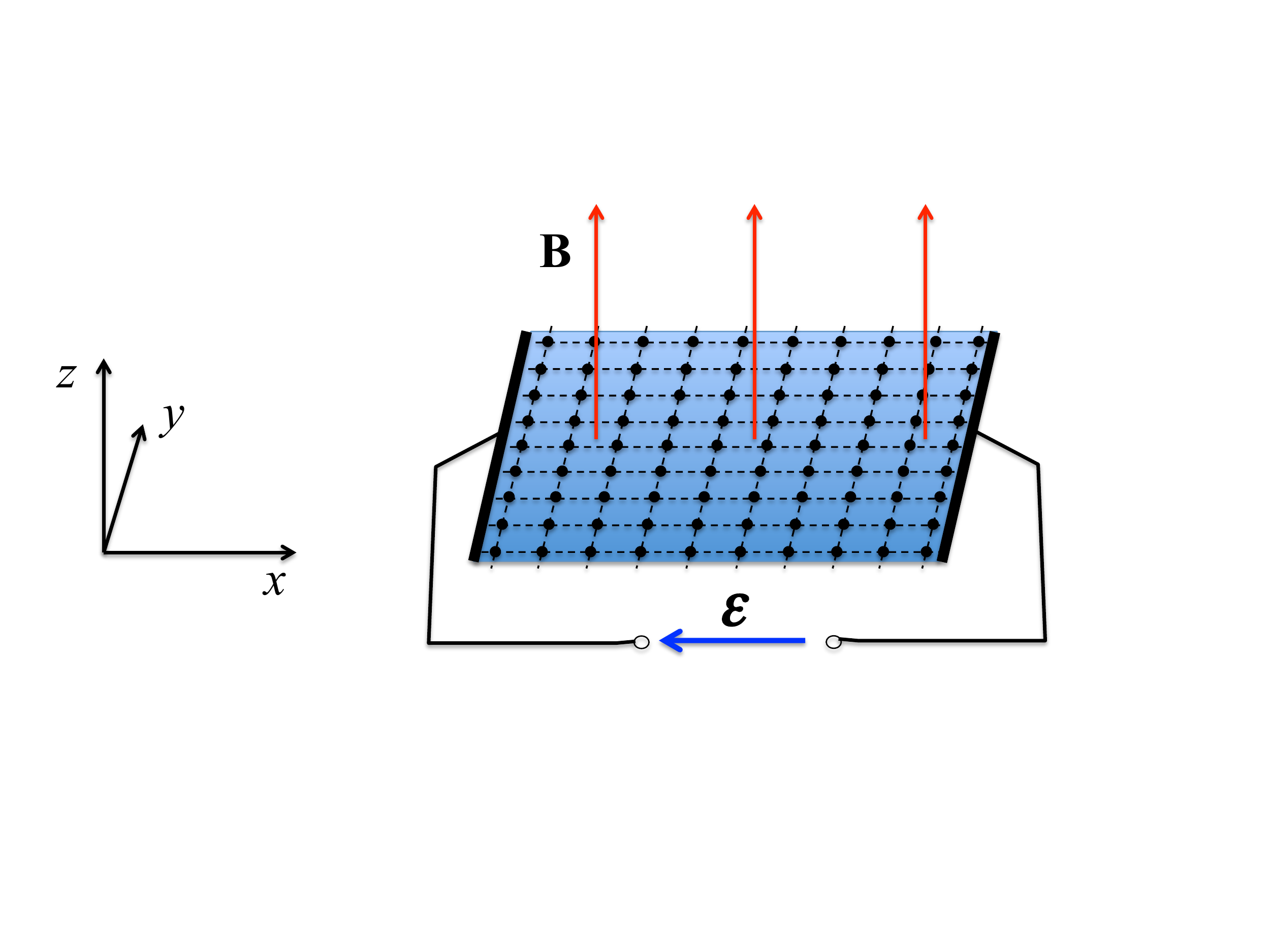}
\vspace{-2.2cm} 
\caption{(Color online) (a) Scheme of a quantum Hall geometry for  
electron transport on a two-dimensional lattice under the simultaneous presence
of static electric and magnetic fields.} \label{fig1}
\end{center}
\end{figure}

\section{Single-electron transport in electric and magnetic fields}

The single-electron hamiltonian in a 2D crystal with a magnetic field applied orthogonally
to the periodicity plane is 
\begin{equation}
\hat{H}=\frac{1}{2\mathrm{m}}\left(\mathbf{p}+\frac{e}{c}\mathbf{A}\right)^{2}+{V}_c (\mathbf{r}) \, ,\label{eq:5}
\end{equation}
where $\mathbf{p}=(p_{x},p_{y})$ is the electron momentum, $\mathbf{A}=(A_{x},A_{y})$ is the
vector potential associated to the applied magnetic field $\mathbf{B} = \nabla \times \mathbf{A}$, ${V}_c (\mathbf{r})={V}_c (x,y)$ 
is the 2D periodic crystal potential, and $\mathrm{m}$ the two-dimensional effective electron mass.  
The real space configuration of a quantum Hall system is schematically represented in Fig.~\ref{fig1}.

In general, the hamiltonian (\ref{eq:5}) lacks translational invariance because of the presence of the 
vector potential, $\mathbf{A}$.
However, this model is still invariant by translational symmetry if the ratio between the magnetic field flux 
entering the original unitary cell and the magnetic flux quantum ($\Phi_{0}={hc}/{e}$) is a rational number \cite{key-21}. 
It is then possible to extend the validity of this condition to values of $\Phi / \Phi_{0}$ arbitrary 
close to any irrational number, with a negligible error \cite{key-23,key-24,key-25}.
Hence, we can always assume that the eigenfunctions of (\ref{eq:5}) are of the Bloch type 
\begin{equation}
\psi_{n\mathbf{k}}(\mathbf{r})=e^{i\mathbf{k}\cdot\mathbf{r}}u_{n\mathbf{k}}(\mathbf{r}) \, ,\label{eq:6}
\end{equation}
and the eigenvalue equation reads
\begin{equation}
\hat{H}\psi_{n\mathbf{k}}(\mathbf{r})=E_{n\mathbf{k}}\psi_{n\mathbf{k}}(\mathbf{r}) \, ,\label{eq:7}
\end{equation}
where $n$ is now interpreted as a {magnetic} band index, $\mathbf{k}$ is still the Bloch
wave vector, and $u_{n\mathbf{k}}(\mathbf{r})$ is the periodic part of the Bloch wave function. 
It is easy to see that using Eq.~(\ref{eq:6}) it is also possible to obtain the parametric eigenvalue equation 
for the $u_{n\mathbf{k}}(\mathbf{r})$, which directly corresponds Eq.~(\ref{eq:7}) and reads
\begin{equation}
\hat{H}_{\mathbf{k}}u_{n\mathbf{k}}(\mathbf{r})=E_{n\mathbf{k}}u_{n\mathbf{k}}(\mathbf{r}) \, .\label{eq:7-1}
\end{equation}

The effects of a static electric field (the Hall field, $\mathbf{\mathcal{E}}$) on the single-particle dynamics 
can be described by using a perturbative approach. 
The perturbed hamiltonian will be a sum of the zero-order hamiltonian, Eq.~(\ref{eq:5}), and a perturbation
term given by 
\begin{equation}
{V}_{{p}}=e\vec{\mathcal{E}}\cdot\mathbf{r} \, .\label{eq:8}
\end{equation}
Up to first order in perturbation theory, the eigenvalues $\tilde{E}_{n}(\mathbf{k})$
and eigenvectors $\bigl|\tilde{\psi}_{n\mathbf{k}}\bigr\rangle$ read
\begin{alignat}{1}
\tilde{E}_{n\mathbf{k}} & \simeq E_{n\mathbf{k}}+\bigl\langle\psi_{n\mathbf{k}}\bigl |{V}_{{p}}\bigr|\psi_{n\mathbf{k}}\bigr\rangle,\label{eq:9}\\
\bigl|\tilde{\psi}_{n\mathbf{k}}\bigr\rangle & \simeq\left|\psi_{n\mathbf{k}}\right\rangle +\sum_{m\neq n}\left|\psi_{m\mathbf{k}}\right\rangle \frac{\bigl\langle\psi_{m\mathbf{k}}\bigl 
| {V}_{{p}}\bigr|\psi_{n\mathbf{k}}\bigr\rangle}{E_{n\mathbf{k}}-E_{m\mathbf{k}}} \, .\label{eq:10}
\end{alignat}
We notice that  there is no mixing in $\mathbf{k}$ (horizontal mixing) in Eq.~(\ref{eq:10}), since 
${V}_{{p}}$ is an electric dipole term with a static electric field,  which does not produce mixing of different states 
within the first Brillouin zone. 
To proceed with the calculation of the conductivity in this system, we first calculate the expectation
value of the group velocity in a perturbed state, within the framework of the Hellmann-Feynman (HF) theorem
\cite{foot2}, whose validity is guaranteed by the fact that the perturbed states 
preserve the Bloch form to first approximation
\begin{alignat}{1}
\bigl|\tilde{\psi}_{n\mathbf{k}}\bigr\rangle & \simeq e^{i\mathbf{k}\cdot\mathbf{r}}\left|u_{n\mathbf{k}}\right\rangle +e^{i\mathbf{k}\cdot\mathbf{r}}\sum_{m\neq n}\left|u_{m\mathbf{k}}\right\rangle \frac{\bigl\langle u_{m\mathbf{k}}\bigl | {V}_{{p}}\bigr|u_{n\mathbf{k}}\bigr\rangle}{E_{n\mathbf{k}}-E_{m\mathbf{k}}},\nonumber \\
 & \simeq e^{i\mathbf{k}\cdot\mathbf{r}}\left|\tilde{u}_{n\mathbf{k}}\right\rangle \, ,
\end{alignat}
where, in the spirit of $\mathbf{k} \cdot \mathbf{p}$ theory, we have defined 
\begin{equation}
\left|\tilde{u}_{n\mathbf{k}}\right\rangle \doteq\left|u_{n\mathbf{k}}\right\rangle +\sum_{m\neq n}\left|u_{m\mathbf{k}}\right\rangle \frac{\bigl\langle u_{m\mathbf{k}}\bigl
|{V}_{p} \bigr |u_{n\mathbf{k}}\bigr\rangle}{E_{n\mathbf{k}}-E_{m\mathbf{k}}} \, .\label{eq:12}
\end{equation}
Redefining for ease of notation $\left|\tilde{u}_{n\mathbf{k}}\right\rangle \doteq\left|\tilde{n}\right\rangle$,
from Eqs.~(\ref{eq:6}), (\ref{eq:7}), and (\ref{eq:7-1}) 
the expectation value for the group velocity of the electron on the state $\bigl|\tilde{\psi}_{n\mathbf{k}}\bigr\rangle$ is
\begin{equation}
\mathbf{\tilde{v}}_{n\mathbf{k}}=\left\langle \tilde{n}\left|\frac{1}{\mathrm{m}}(\mathbf{p}+\hslash\mathbf{k})\right|\tilde{n}\right\rangle =
\left\langle \tilde{n}\left|\nabla_{\mathbf{k}}\hat{H}_{\mathbf{k}}\right|\tilde{n}\right\rangle \,  ,\label{eq:13}
\end{equation}
from which, using Eq.~(\ref{eq:12}), we get (neglecting higher order terms)
\begin{alignat}{1} 
\mathbf{\tilde{v}}_{n\mathbf{k}} \simeq  & \left\langle n\left| \frac{1}{\mathrm{m}}(\mathbf{p}+\hslash\mathbf{k}) \right| n \right\rangle +\nonumber \\
 & +\sum_{m\neq n} \left\langle m\left| \frac{1}{\mathrm{m}}(\mathbf{p}+\hslash\mathbf{k}) \right| n\right\rangle \frac{\left\langle n\right|e\vec{\mathcal{E}}\cdot\mathbf{r}\left|m\right\rangle }{E_{n\mathbf{k}}-E_{m\mathbf{k}}}+\nonumber \\
 & +\sum_{m'\neq n}\left\langle n\left| \frac{1}{\mathrm{m}}(\mathbf{p}+\hslash\mathbf{k})\right| m'\right\rangle \frac{\left\langle m'\right|e\vec{\mathcal{E}}\cdot\mathbf{r}\left|n\right\rangle }{E_{n\mathbf{k}}-E_{m'\mathbf{k}}} \, . \label{eq:14} 
\end{alignat}
With this notation, the HF equations read
\begin{equation}
\left\langle m\left|\frac{1}{\mathrm{m}}(\mathbf{p}+\hslash\mathbf{k}) \right| n\right\rangle 
= \frac{E_{n\mathbf{k}}-E_{m\mathbf{k}}}{\hslash}\left\langle m|\nabla_{\mathbf{k}}n\right\rangle ,\label{eq:16}
\end{equation}
\begin{equation}
\left\langle m\right|\mathbf{r}\left|n\right\rangle =i\left\langle m|\nabla_{\mathbf{k}}n\right\rangle \, ,\label{eq:17}
\end{equation}
valid for $m\neq n$. 
By using Eq.~(\ref{eq:16}) in Eq.~(\ref{eq:14}), and assuming (without loss of generality) that the Hall
field is directed along $x$, $\vec{\mathcal{E}}=\mathcal{E}_H \hat{\mathbf{x}}$ ($\hat{\mathbf{x}}$ 
indicates the unit vector in the $x$ direction), we have
\begin{alignat}{1}
\mathbf{\tilde{v}}_{n\mathbf{k}} & =\frac{1}{\hslash}\nabla_{\mathbf{k}}E_{n\mathbf{k}}+ \nonumber \\
 & +\sum_{m'\neq n}\sum_{m\neq n}\frac{ie\mathcal{E}_H}{\hslash}\left(\left\langle m|\nabla_{\mathbf{k}}n\right\rangle \left\langle n\right|x\left|m\right\rangle +\right.\nonumber \\
 & \left.-\left\langle n|\nabla_{\mathbf{k}}m'\right\rangle \left\langle m'\right|x\left|n\right\rangle \right) \, , \label{eq:19}  
\end{alignat}
from which, using Eq.~(\ref{eq:17}) we get
\begin{alignat}{1}
\mathbf{\tilde{v}}_{n\mathbf{k}} & =\frac{1}{\hslash}\nabla_{\mathbf{k}}E_{n\mathbf{k}}+\nonumber\\
 & +\sum_{m'\neq n}\sum_{m\neq n}\frac{ie\mathcal{E}_H}{\hslash}\left(\left\langle m|\nabla_{\mathbf{k}}n\right\rangle \left\langle n|\partial_{k_{x}}m\right\rangle \right.\nonumber \\
 & \left.-\left\langle n|\nabla_{\mathbf{k}}m'\right\rangle \left\langle m'|\partial_{k_{x}}n\right\rangle \right) \, . \label{eq:20} 
\end{alignat}
From Eq.~(\ref{eq:17}) it is straightforward to show that
\begin{equation}
i\left\langle n|\nabla_{\mathbf{k}}m\right\rangle =-i\left\langle \nabla_{\mathbf{k}}n|m\right\rangle \, ,\label{eq:21}
\end{equation}
which we can plug in Eq.~(\ref{eq:20}) to obtain  
\begin{alignat}{1}
\mathbf{\tilde{v}}_{n\mathbf{k}} & =\frac{1}{\hslash}\nabla_{\mathbf{k}}E_{n\mathbf{k}}+ \nonumber \\
 & -\sum_{m'\neq n}\sum_{m\neq n}\frac{ie\mathcal{E}_H}{\hslash}\left[ \left\langle \partial_{k_{x}}n |m \right\rangle \left\langle m|\nabla_{\mathbf{k}}n\right\rangle +\right.\nonumber \\
 & \left.-\left\langle \nabla_{\mathbf{k}}n|m'\right\rangle \left\langle m'|\partial_{k_{x}}n\right\rangle \right] \, . \label{eq:22} 
\end{alignat}
By using the completeness relation
\begin{equation}
\sum_{m}\left|m\right\rangle \left\langle m\right|=\sum_{m\neq n}\left|m\right\rangle \left\langle m\right|+\left|n\right\rangle \left\langle n\right|=\mathbb{I} \, ,\label{eq:23}
\end{equation}
Eq.~(\ref{eq:22}) is reduced to
\begin{alignat}{1}
\mathbf{\tilde{v}}_{n\mathbf{k}}= & \frac{1}{\hslash}\nabla_{\mathbf{k}}E_{n\mathbf{k}}+ \nonumber \\
 & \frac{ie\mathcal{E}_H \hat{\mathbf{y}}}{\hslash}\left[\left\langle \partial_{k_{x}}n|\partial_{k_{y}}n\right\rangle -\left\langle \partial_{k_{y}}n|\partial_{k_{x}}n\right\rangle \right] \, ,\label{eq:24} 
\end{alignat}
where we have indicated with $\hat{\mathbf{y}}$ the unit vector in the $y$ direction.
The last equation can be rewritten as 
\begin{alignat}{1}
\label{eq:25}\\
\mathbf{\tilde{v}}_{n\mathbf{k}} =
%\frac{1}{\hslash}\nabla_{\mathbf{k}}E_{n\mathbf{k}}-\frac{ie\mathcal{\left|E\right|}\hat{\mathbf{y}} }{\hslash}\left[\left\langle \nabla_{\mathbf{k}}n|\times|\nabla_{\mathbf{k}}n\right\rangle \cdot\hat{\mathbf{z}}\right]\nonumber \\
 \frac{1}{\hslash}\nabla_{\mathbf{k}}E_{n\mathbf{k}}-\frac{ie\mathcal{E}_H\hat{\mathbf{y}}}{\hslash}\left[\left(rot\left\langle n|\nabla_{\mathbf{k}}n\right\rangle \right)\cdot\hat{\mathbf{z}}\right] \, ,\nonumber 
\end{alignat}
and defining the Berry curvature $\boldsymbol{\Omega}_{n\mathbf{k}}\doteq i\left(rot\left\langle n|\nabla_{\mathbf{k}}n\right\rangle \right)$ we finally get
\begin{equation}
\mathbf{\tilde{v}}_{n\mathbf{k}}=\frac{1}{\hslash}\nabla_{\mathbf{k}}E_{n\mathbf{k}}+\frac{e}{\hslash}\vec{\mathcal{E}}\times\boldsymbol{\Omega}_{n\mathbf{k}} \, , \label{eq:26-1}
\end{equation}
which can be recast in the more familiar and well known expression \cite{key-17,key-18,key-19}, Eq.~(\ref{eq:2}), 
by using the semiclassical equation of motion, Eq.~(\ref{eq:3}).

The importance of the topological term defined by the Berry curvature in Eq.~(\ref{eq:26-1}) emerges 
after calculating the Hall conductance for this system. 
We briefly report here this calculation, for the sake of completeness. In the simplest thermodynamical case in which the temperature 
of the system is $T=0$ K, the contribution of a given magnetic Bloch band to the drift velocity is
\begin{alignat}{1}
\mathbf{v}_{\mathrm{d},n} & =\frac{V}{4\pi^{2}}\intop_{B.Z.} \mathrm{d}^2 \mathbf{k} \,\, \mathbf{\tilde{v}}_{n\mathbf{k}} % \nonumber   \\
% & =-\frac{V}{4\pi^{2}}\intop_{B.Z.}d\mathbf{k}^{2}\dot{\mathbf{k}}\times\boldsymbol{\Omega}_{n\mathbf{k}}\nonumber \\
 =\frac{V}{4\pi^{2}}\frac{e\mathcal{E}_H \hat{\mathbf{y}} }{\hslash}\intop_{B.Z.} \mathrm{d}^2 \mathbf{k} \left(\boldsymbol{\Omega}_{n\mathbf{k}}\cdot\hat{\mathbf{z}}\right) \, ,
 \label{eq:27} 
\end{alignat}
where $V$ is the volume of the primitive cell, $\boldsymbol{\Omega}_{n\mathbf{k}}\cdot\hat{\mathbf{z}}$ is the component of the Berry curvature along $z$, and the integral is performed over the first Brillouin zone for which, using Eq.~(\ref{eq:2}), the term $\nabla_{\mathbf{k}}E_{n\mathbf{k}}$ does not contribute. 
From the last equation we straightforwardly get the current density contributed by the given band
\begin{equation}
\mathbf{J}_{n}=-\frac{1}{4\pi^{2}}\frac{e^{2}\mathcal{E}_H \hat{\mathbf{y}}}{\hslash}\intop_{B.Z.} \mathrm{d}^2\mathbf{k} \left(\boldsymbol{\Omega}_{n\mathbf{k}}\cdot\hat{\mathbf{z}}\right) \, ,\label{eq:28}
\end{equation}
from which the transverse conductivity (in 2D, the Hall conductance) is quantized and given by integer multiples of the quantum of conductance, $e^2 /h$, as
\begin{equation}
\sigma_{n}^{xy}=-\frac{1}{4\pi^{2}}\frac{e^{2}}{\hslash}\intop_{B.Z.}\mathrm{d}^2\mathbf{k} \left(\boldsymbol{\Omega}_{n\mathbf{k}}\cdot\hat{\mathbf{z}}\right)=-\frac{e^{2}}{h}\mathcal{C}_{n} \, ,\label{eq:29}
\end{equation}
where $\mathcal{C}_{n}=\frac{1}{2\pi}\intop_{B.Z.} \mathrm{d}^2\mathbf{k} \left(\boldsymbol{\Omega}_{n\mathbf{k}}\cdot\hat{\mathbf{z}}\right)$
is exactly the well known expression for the Chern number \cite{key-16,simon83}, which we have independently obtained here.

\section{Theory of photonic transport in gyrotropic 2D photonic crystals}

Time-reversal symmetry (TRS) breaking is responsible for the topological nature of the integer quantum Hall
phenomenology, which is a strong indication that an analogous effect must exist in photonic 
band gap media with broken TRS, as pointed out by Haldane and Raghu \cite{key-13,key-5}.
In order to rigorously check the deep connections between electronic and
photonic semiclassical dynamics, we hereby develop a bulk topological theory
for weakly perturbed photonic crystals with broken TRS,
which will lead to an equation for the velocity of the electromagnetic mode containing 
a topological term formally equivalent to Eq.~(\ref{eq:2}), thus enforcing the analogies
between Schr\"{o}dinger and Maxwell equations. 

In the most general case where the dielectric permittivity $\overset{\leftrightarrow}{\boldsymbol{\varepsilon}}$
and the magnetic permeability $\overset{\leftrightarrow}{\boldsymbol{\mu}}$
are second-order tensors, Maxwell equations in photonic crystals can be written in the form of a generalized 
eigenvalue problem  (see, e.g., Ref.~\onlinecite{key-14})
\begin{flalign}
\boldsymbol{\Pi}_{{e}}\mathbf{E}\left(\mathbf{r}\right) 
& =\omega^{2}\overset{\leftrightarrow}{\boldsymbol{\varepsilon}}(\mathbf{r})\mathbf{E}\left(\mathbf{r}\right)\label{eq:30}\\
\boldsymbol{\Pi}_{{m}}\mathbf{H}\left(\mathbf{r}\right) 
& =\omega^{2}\overset{\leftrightarrow}{\boldsymbol{\mu}}(\mathbf{r})\mathbf{H}\left(\mathbf{r}\right)\, ,\label{eq:31}
\end{flalign}
where $\mathbf{E}\left(\mathbf{r}\right)$ and $\mathbf{H}\left(\mathbf{r}\right)$ are the electric and the magnetic 
fields, respectively, and $\omega$ is the oscillation frequency, while 
\begin{gather}
\boldsymbol{\Pi}_{{e}}\doteq\nabla\times\left(\overset{\leftrightarrow}{\boldsymbol{\mu}}^{-1}(\mathbf{r})\nabla\times\bullet\right) \, \label{eq:32}\\
\boldsymbol{\Pi}_{{m}}\doteq\nabla\times\left(\overset{\leftrightarrow}{\boldsymbol{\varepsilon}}^{-1}(\mathbf{r})\nabla\times\bullet\right) \, \label{eq:33}
\end{gather}
are linear operators of the generalized eigenvalue problem.
Such eigenvalue problem can be recast in a standard one by using the following basis states \cite{foot3,key-27,key-28}
\begin{gather}
\boldsymbol{F}_{{e}}(\mathbf{r})=\overset{\leftrightarrow}{\boldsymbol{\varepsilon}}^{\frac{1}{2}}(\mathbf{r})\mathbf{E}(\mathbf{r})\label{eq:34}\\
\boldsymbol{F}_{{m}}(\mathbf{r})=\overset{\leftrightarrow}{\boldsymbol{\mu}}^{\frac{1}{2}}(\mathbf{r})\mathbf{H}(\mathbf{r}) \, ,\label{eq:35}
\end{gather}
which allow to obtain the eigenvalue equations
\begin{gather}
\boldsymbol{\Theta}_{{e}}\boldsymbol{F}_{{e}}(\mathbf{r})=\omega^{2}\boldsymbol{F}_{{e}}(\mathbf{r}) \, \label{eq:36}\\
\boldsymbol{\Theta}_{{m}}\boldsymbol{F}_{{m}}(\mathbf{r})=\omega^{2}\boldsymbol{F}_{{m}}(\mathbf{r}) \, ,\label{eq:37}
\end{gather}
where the hermitian operators are defined as 
\begin{alignat}{1}
\boldsymbol{\Theta}_{{e}}\doteq\overset{\leftrightarrow}{\boldsymbol{\varepsilon}}^{-\frac{1}{2}}(\mathbf{r})\nabla\times\left[\overset{\leftrightarrow}{\boldsymbol{\mu}}^{-1}(\mathbf{r})\nabla\times\left(\overset{\leftrightarrow}{\boldsymbol{\varepsilon}}^{-\frac{1}{2}}(\mathbf{r})\bullet\right)\right]\label{eq:38}\\
\boldsymbol{\Theta}_{{m}}\doteq\overset{\leftrightarrow}{\boldsymbol{\mu}}^{-\frac{1}{2}}(\mathbf{r})\nabla\times\left[\overset{\leftrightarrow}{\boldsymbol{\varepsilon}}^{-1}(\mathbf{r})\nabla\times\left(\overset{\leftrightarrow}{\boldsymbol{\mu}}^{-\frac{1}{2}}(\mathbf{r})\bullet\right)\right] \, .\label{eq:39}
\end{alignat}
{Normalization of the fields is well defined by the notion of scalar product, $\langle \mathbf{F}_{e,m} | \mathbf{F}_{e,m} \rangle = \int \mathrm{d}^3 \mathbf{r} \, \mathbf{F}^{\ast}_{e,m} (\mathbf{r})\mathbf{F}_{e,m} (\mathbf{r})$ and the physical requirement that the electromagnetic energy density be finite in the system \cite{key-14}.}
Since the two eigenvalue equations are perfectly specular with each other, we will focus on the equation for the 
electric field henceforth. 
Following the proposal in \cite{key-11}, we allow TRS breaking in the system by using 2D gyrotropic 
photonic crystals. 
For practical purposes, we assume a 2D square lattice of YIG (Yttrium iron garnet) rods in air \cite{key-11},
without loss of generality of the formalism.
The permittivity and the permeability of this system can be explicitly represented as 
\begin{equation}
\overset{\leftrightarrow}{\boldsymbol{\varepsilon}}=\left[\begin{array}{ccc}
\varepsilon(\mathbf{r}) & 0 & 0\\
0 & \varepsilon(\mathbf{r}) & 0\\
0 & 0 & \varepsilon(\mathbf{r})
\end{array}\right],\label{eq:40}
\end{equation}
\begin{equation}
\overset{\leftrightarrow}{\boldsymbol{\mu}}=\left[\begin{array}{ccc}
\mu(\mathbf{r}) & i\gamma(\mathbf{r}) & 0\\
-i\gamma(\mathbf{r}) & \mu(\mathbf{r}) & 0\\
0 & 0 & \mu_{0}
\end{array}\right] \, ,\label{eq:41}
\end{equation}
and the inverse of $\overset{\leftrightarrow}{\boldsymbol{\mu}}$ is 
\begin{equation}
\overset{\leftrightarrow}{\boldsymbol{\mu}}^{-1}=\left[\begin{array}{ccc}
\bar{\mu}^{-1}(\mathbf{r}) & i\eta(\mathbf{r}) & 0\\
-i\eta(\mathbf{r}) & \bar{\mu}^{-1}(\mathbf{r}) & 0\\
0 & 0 & \mu_{0}^{-1}
\end{array}\right] \, ,\label{eq:42}
\end{equation}
where $\bar{\mu}^{-1}(\mathbf{r})\doteq\frac{\mu(\mathbf{r})}{\mu^{2}(\mathbf{r})-\gamma^{2}(\mathbf{r})}$,
and  $\eta(\mathbf{r})\doteq\frac{-\gamma(\mathbf{r})}{\mu^{2}(\mathbf{r})-\gamma^{2}(\mathbf{r})}$.

{In the following, and in full analogy to the electron dynamics reported above, we will assume a 2D photon dynamics, where mirror symmetry with respect to the propagation plane allows to define even (transverse-electric, TE) and odd (transverse-magnetic, TM) modes, respectively \cite{key-14}. }
Moreover, as it can be seen from Eq.~(\ref{eq:42}), we have introduced a magnetic ``Faraday mixing'' only in the $xy$ plane, which means that we can restrict our analysis to the 
TM modes only, i.e.  $(\mathrm{H_{x}},\mathrm{H_{y}},\mathrm{E_{z}})$ field components different from zero. {This assumption is realistic for the cases usually considered for 2D photonic crystals with gyrotropic constituents (see also discussion in Sec.~\ref{sec:discussion}), where no mixing of the two polarization eigenstates occurs. }
An eigenvalue equation for the scalar problem is then explicitly derived as (see App.~A for the derivation details)
\begin{flalign}
\boldsymbol{\Theta}F_{z} = \omega^{2}F_{z} \, ,  \label{eq:43} 
\end{flalign}
where $F_{z}$ is the $z$ component of the vector $\boldsymbol{F}$ (we have dropped the subscript $e$ for easier notation), and the operator is explicitly given by
\begin{flalign}
\boldsymbol{\Theta} & =\biggl[-\varepsilon^{-1}\bar{\mu}^{-1}\nabla^{2}+ \nonumber \\
 & -\left(\bar{\mu}^{-1}\nabla\varepsilon^{-1}+\varepsilon^{-1}\nabla\bar{\mu}^{-1}+i\varepsilon^{-1}\left(\hat{\mathbf{z}} \times\nabla\eta\right)\right)\cdot\nabla+\nonumber \\
 & -\frac{1}{2}\nabla\bar{\mu}^{-1}\cdot\nabla\varepsilon^{-1}-\frac{1}{2}\bar{\mu}^{-1}\nabla^{2}\varepsilon^{-1}+\nonumber \\
 & \left.+\frac{1}{4}\bar{\mu}^{-1}\varepsilon\left(\nabla\varepsilon^{-1}\right)^{2}-\frac{1}{2}i\left(\hat{\mathbf{z}} 
 \times\nabla\eta\right)\cdot\nabla\varepsilon^{-1}\right] \, .  \label{eq:43bis} 
\end{flalign} 
The operator in Eq.~(\ref{eq:43bis}) has translational  symmetry, so its eigenvectors satisfy the Bloch-Floquet theorem \cite{key-14}, and are given by an expression similar to Eq.~(\ref{eq:6}). 
We can rewrite the eigenvalue problem for the periodic part of $F_z$, which we define $u_{n \mathbf{k}}$ to keep the analogy with the electronic case, as in Eq.~(\ref{eq:7-1})
\begin{alignat}{1}
\boldsymbol{\Theta}_{\mathbf{k}}u_{n\mathbf{k}}= & \omega_{n\mathbf{k}}^{2}u_{n\mathbf{k}} \, .\label{eq:44}
\end{alignat}

In order to apply the perturbative approach described in the previous section, we introduce a photonic perturbation mimicking
the effect of an electric field as a dragging force, which is achieved by adding a weak modulation $\Delta\varepsilon$ to the 
periodic permittivity, imposing the following conditions:
\begin{enumerate}
\item $\frac{\Delta\varepsilon}{\varepsilon}\ll1$;
\item $\frac{\Delta\varepsilon}{\varepsilon}$ is slowly varying on the scale determined by the lattice constant, $a$;
\item $\frac{\Delta\varepsilon}{\varepsilon}$ is a linear function of $x$.
\end{enumerate}
We notice that the $x$ axis is chosen here just to preserve the connection with the treatment
given for the electron dynamics in Sec.~II.
As an explicit example and without loss of generality, we can assume $\frac{\Delta\varepsilon}{\varepsilon}=\lambda\frac{x}{a}$, where
$\lambda$ is a small constant. With this slow grading of the permittivity, the perturbed operator
$\tilde{\boldsymbol{\Theta}}$ takes the form (see App.~B for the explicit derivation)
\begin{alignat}{1}
\tilde{\boldsymbol{\Theta}} & =\boldsymbol{\Theta}-\lambda\frac{x}{a}\boldsymbol{\Theta} \nonumber \\
 & =\boldsymbol{\Theta}+\mathbf{V}^{p} \, , \label{eq:45}  
\end{alignat}
where we have implicitly defined 
\begin{alignat}{1}
\mathbf{V}^{p} & \doteq-\lambda\frac{x}{a}\boldsymbol{\Theta} \, .\label{eq:46}
\end{alignat}
Using the perturbation theory up to the first order, we get the perturbed eigenvectors and eigenstates
\begin{alignat}{1}
\tilde{\omega}_{n\mathbf{k}}^{2} & \simeq\omega_{n\mathbf{k}}^{2}+\left\langle F_{n\mathbf{k}}|\boldsymbol{V}^{p} F_{n\mathbf{k}}\right\rangle \label{eq:47}\\
\bigl|\tilde{F}_{n\mathbf{k}}\bigr\rangle & \simeq\left| F_{n\mathbf{k}}\right\rangle +\sum_{m\neq n}\left|F_{m\mathbf{k}}\right\rangle \frac{\left\langle F_{m\mathbf{k}}|\boldsymbol{V}^{p} F_{n\mathbf{k}}\right\rangle }{\omega_{n\mathbf{k}}^{2}-\omega_{m\mathbf{k}}^{2}} \, ,\label{eq:48}
\end{alignat}
where by $F_{n \mathbf{k}}$ we mean the $z$-component of the Bloch eigenfunction, $\mathbf{F}$. Exactly as done in the previous section we then write
\begin{alignat}{1}
\left|\tilde{u}_{n\mathbf{k}}\right\rangle  & \doteq\left|u_{n\mathbf{k}}\right\rangle +\sum_{m\neq n}\left|u_{m\mathbf{k}}\right\rangle \frac{\left\langle u_{m\mathbf{k}}|\boldsymbol{V}_{\mathbf{k}}^{p} u_{n\mathbf{k}}\right\rangle }{\omega_{n\mathbf{k}}^{2}-\omega_{m\mathbf{k}}^{2}}\label{eq:49}\\
\boldsymbol{V}_{\mathbf{k}}^{p} & \doteq e^{-i\mathbf{k}\cdot\mathbf{r}}\boldsymbol{V}^{p}e^{i\mathbf{k}\cdot\mathbf{r}} \, .\label{eq:50}
\end{alignat}
{In this framework, we notice that we are conceptually exploiting an adaptation of the $\mathbf{k} \cdot \mathbf{p}$ theory \cite{johnson,sipe2000,busch}. To avoid mathematical issues at degeneracy points in the first Brillouin zone, we are assuming non-degenerate photonic bands here throughout the manuscript.}

The dynamical properties will be given by calculating the electromagnetic field velocity.
However, a note of warning is worth here.  In fact, while the physical velocity of an electromagnetic mode, 
i.e. the one associated to the  electromagnetic energy flux from the Poynting vector, 
$\mathbf{S}_{n\mathbf{k}}=\frac{1}{2}\mathrm{Re}\{ \mathbf{E}_{n\mathbf{k}}^{*}\times\mathbf{H}_{n\mathbf{k}} \}$, is 
given by
\begin{alignat}{1} 
\mathit{\mathbf{v}}_{n\mathbf{k}}^{(e)}  =
\frac{\int \mathrm{d}^{3}\mathbf{r} \, \, \mathbf{S}_{n\mathbf{k}}}{\mathcal{U}_{n\mathbf{k}} } \, ,\label{eq:51}
\end{alignat}
where the electromagnetic energy density is expressed as $\mathcal{U}_{n\mathbf{k}} =  \mathcal{U}_{n\mathbf{k}}^{{e}}+\mathcal{U}_{n\mathbf{k}}^{{m}}$, with
$\mathcal{U}_{n\mathbf{k}}^{{e}} = \frac{1}{4}\int \mathrm{d}^{3}\mathbf{r} (\overset{\leftrightarrow}{\boldsymbol{\varepsilon}}\mathbf{E}_{n\mathbf{k}})\cdot  \mathbf{E}_{n\mathbf{k}}^{*}$ and $\mathcal{U}_{n\mathbf{k}}^{{m}} = \frac{1}{4}\int \mathrm{d}^{3}\mathbf{r} (\overset{\leftrightarrow} {\boldsymbol{\mu}}\mathbf{H}_{n\mathbf{k}})\cdot  \mathbf{H}_{n\mathbf{k}}^{*}$,  the group velocity of the mode is actually given by 
\begin{alignat}{1}
\mathit{\mathbf{v}}_{n\mathbf{k}}^{(g)} & =\nabla_{\mathbf{k}} \omega 
 =\frac{1}{2\omega_{n\mathbf{k}}}\frac{\bigl\langle\boldsymbol{u}_{n\mathbf{k}}\bigr|\nabla_{\mathbf{k}}\boldsymbol{\Theta}_{\mathbf{k}}\bigl|\boldsymbol{u}_{n\mathbf{k}}\bigr\rangle}{\bigl\langle\boldsymbol{u}_{n\mathbf{k}}\bigl|\boldsymbol{u}_{n\mathbf{k}}\bigr\rangle} \, , \label{eq:52}
\end{alignat}
{where the last equality is the photonic crystal version of Eq.~(\ref{eq:13}), as in Ref.~\onlinecite{key-14} (see also App.~C). } 
In an ideal photonic crystal {made of non-dispersive constituents}, one can show that $\mathit{\mathbf{v}}_{n\mathbf{k}}^{(e)} = \mathit{\mathbf{v}}_{n\mathbf{k}}^{(g)}$ \cite{key-14,key-29}, {as it has been specifically shown for generic 2D photonic crystals in a photonic $\mathbf{k} \cdot \mathbf{p}$ framework \cite{foteinopoulou2005}. }
Even if the equality between energy and group velocity is not generally fulfilled in perturbed systems, {it can be shown (see App.~D) that in the case of non-dispersive media (i.e., for frequency-independent permittivity and permeability tensors)} the energy velocity of the mode can be defined as 
\begin{alignat}{1}
\tilde{\mathit{\mathbf{v}}}_{n\mathbf{k}}^{(e)}= & \frac{1}{2\omega_{n\mathbf{k}}}\frac{\left\langle \tilde{u}_{n\mathbf{k}}\left|\nabla_{\mathbf{k}}\boldsymbol{\Theta}_{\mathbf{k}}\right|\tilde{u}_{n\mathbf{k}}\right\rangle }{\left\langle {u}_{n\mathbf{k}}| {u}_{n\mathbf{k}}\right\rangle } \, ,\label{eq:53}
\end{alignat}
where we are implicitly assuming that, up to first order in perturbation theory, we can approximate $\left\langle \tilde{u}_{n\mathbf{k}}| \tilde{u}_{n\mathbf{k}}\right\rangle \simeq \left\langle {u}_{n\mathbf{k}}| {u}_{n\mathbf{k}}\right\rangle $ in the denominator (as we have done throughout App.~D).

Using now Eq.~(\ref{eq:49}), and redefining $\left|\tilde{u}_{n\mathbf{k}}\right\rangle \doteq\left|\tilde{n}\right\rangle $
and $\left|u_{n\mathbf{k}}\right\rangle \doteq\left|n\right\rangle $ for ease of notation, Eq.~(\ref{eq:53}) can be written as
\begin{alignat}{1}
\tilde{\mathit{\mathbf{v}}}_{n\mathbf{k}}^{(e)}= & \frac{1}{2\omega_{n\mathbf{k}}\left\langle {n}| {n}\right\rangle }\cdot  \nonumber \\
 & \cdot\left[\left(\left\langle n\right|+\sum_{m\neq n}\left\langle m\right|\frac{\left\langle \boldsymbol{V}_{\mathbf{k}}^{p}n|m\right\rangle }{\omega_{n\mathbf{k}}^{2}-\omega_{m\mathbf{k}}^{2}}\right)\nabla_{\mathbf{k}}\boldsymbol{\Theta}_{\mathbf{k}}\cdot\right.\nonumber \\
 & \left.\cdot\left(\left|n\right\rangle +\sum_{m'\neq n}\left|m'\right\rangle \frac{\left\langle m'|\boldsymbol{V}_{\mathbf{k}}^{p}n\right\rangle }{\omega_{n\mathbf{k}}^{2}-\omega_{m'\mathbf{k}}^{2}}\right)\right] \, , \label{eq:54} 
\end{alignat}
from which, taking into account only the first order terms, we get
\begin{alignat}{1}
\tilde{\mathit{\mathbf{v}}}_{n\mathbf{k}}^{(e)}= & \frac{1}{2\omega_{n\mathbf{k}}\left\langle n|n\right\rangle }\Biggl[\left\langle n\left|\nabla_{\mathbf{k}}\boldsymbol{\Theta}_{\mathbf{k}}\right|n\right\rangle + \nonumber \\
 & +\sum_{m'\neq n}\left\langle n\left|\nabla_{\mathbf{k}}\boldsymbol{\Theta}_{\mathbf{k}}\right|m'\right\rangle \frac{\left\langle m'|\boldsymbol{V}_{\mathbf{k}}^{p}n\right\rangle }{\omega_{n\mathbf{k}}^{2}-\omega_{m'\mathbf{k}}^{2}}+\nonumber \\
 & +\sum_{m\neq n}\left\langle m\left|\nabla_{\mathbf{k}}\boldsymbol{\Theta}_{\mathbf{k}}\right|n\right\rangle \frac{\left\langle \boldsymbol{V}_{\mathbf{k}}^{p}n|m\right\rangle }{\omega_{n\mathbf{k}}^{2}-\omega_{m\mathbf{k}}^{2}}\Biggr] \, . \label{eq:55}  
\end{alignat}
Using now the first of the photonic HF equations (see App.~C for details)
\begin{equation}
\bigl\langle\boldsymbol{u}_{m\mathbf{k}}\bigr|\nabla_{\mathbf{k}}\boldsymbol{\Theta}_{\mathbf{k}}\bigl|\boldsymbol{u}_{n\mathbf{k}}\bigr\rangle=\left(\omega_{n\mathbf{k}}^{2}-\omega_{m\mathbf{k}}^{2}\right)\bigl\langle\boldsymbol{u}_{n\mathbf{k}}\bigr|\nabla_{\mathbf{k}}\boldsymbol{u}_{n\mathbf{k}}\bigr\rangle \, , \label{eq:56}
\end{equation}
Eq.~(\ref{eq:55}) takes the form
\begin{alignat}{1}
\tilde{\mathit{\mathbf{v}}}_{n\mathbf{k}}^{e} & =\frac{1}{2\omega_{n\mathbf{k}}\left\langle n|n\right\rangle }\biggl[\left\langle n\left|\nabla_{\mathbf{k}}\boldsymbol{\Theta}_{\mathbf{k}}\right|n\right\rangle + \nonumber \\
 & -\sum_{m'\neq n}\left\langle n|\nabla_{\mathbf{k}}m'\right\rangle \left\langle m'|\boldsymbol{V}_{\mathbf{k}}^{p}n\right\rangle +\nonumber \\
 & +\sum_{m\neq n}\left\langle m|\nabla_{\mathbf{k}}n\right\rangle \left\langle n|\boldsymbol{V}_{\mathbf{k}}^{p\boldsymbol{\dagger}}m\right\rangle \biggr] \, . 
 \label{eq:57} 
\end{alignat}
From Eq.~(\ref{eq:46}), we can write that
\begin{alignat}{1}
\left\langle m|\boldsymbol{V}_{\mathbf{k}}^{p}n\right\rangle  & =-\left\langle m|x\frac{\lambda}{a}\boldsymbol{\Theta}_{\mathbf{k}}n\right\rangle \nonumber \\
 & =\sum_{s}-\left\langle m\left|x\right|s\right\rangle \left\langle s|\frac{\lambda}{a}\boldsymbol{\Theta}_{\mathbf{k}}n\right\rangle \nonumber \\
 & =\sum_{s}-\frac{\lambda}{a}\left\langle m\left|x\right|s\right\rangle \omega_{n\mathbf{k}}^{2}\delta_{sn}\nonumber \\
 & =-\frac{\lambda}{a}\left\langle m\left|x\right|n\right\rangle \omega_{n\mathbf{k}}^{2} \, , \label{eq:58}  
\end{alignat}
and using what we have shown in Eq.~(\ref{eq:57}), we get 
\begin{alignat}{1}
\tilde{\mathit{\mathbf{v}}}_{n\mathbf{k}}^{e}= & \frac{1}{2\omega_{n\mathbf{k}}\left\langle n|n\right\rangle }\Biggl[\left\langle n\left|\nabla_{\mathbf{k}}\boldsymbol{\Theta}_{\mathbf{k}}\right|n\right\rangle + \nonumber  \\
 & +\sum_{m'\neq n}\frac{\lambda}{a}\left\langle n|\nabla_{\mathbf{k}}m'\right\rangle \left\langle m'\left|x\right|n\right\rangle \omega_{n\mathbf{k}}^{2}+\nonumber \\
 & -\sum_{m\neq n}\frac{\lambda}{a}\left\langle m|\nabla_{\mathbf{k}}n\right\rangle \left\langle n\left|x\right|m\right\rangle \omega_{n\mathbf{k}}^{2}\Biggr] \, .
 \label{eq:59}  
\end{alignat}
Using now the second photonic HF equation, and the fact that the operator $-i\nabla_{\mathbf{k}}$ is self-adjoint, 
Eq.~(\ref{eq:59}) becomes
\begin{alignat}{1}
\tilde{\mathit{\mathbf{v}}}_{n\mathbf{k}}^{e}= & \frac{1}{2\omega_{n\mathbf{k}}\left\langle n|n\right\rangle }\Biggl[\left\langle n\left| \nabla_{\mathbf{k}}\boldsymbol{\Theta}_{\mathbf{k}}\right|n\right\rangle +  \nonumber  \\
 & -\sum_{m'\neq n}i\frac{\lambda}{a}\omega_{n\mathbf{k}}^{2}\left\langle \nabla_{\mathbf{k}}n|m'\right\rangle \left\langle m'| \partial_{k_{x}}n\right\rangle +\nonumber \\
 & +\sum_{m\neq n}i\frac{\lambda}{a}\omega_{n\mathbf{k}}^{2}\left\langle m|\nabla_{\mathbf{k}}n\right\rangle \left\langle \partial_{k_{x}}n|m\right\rangle \Biggr] \, , \label{eq:61}
\end{alignat}
from which exploiting the completeness relation we straightforwardly obtain
\begin{alignat}{1}
\tilde{\mathit{\mathbf{v}}}_{n\mathbf{k}}^{e}= & \frac{1}{2\omega_{n\mathbf{k}}\left\langle n|n\right\rangle }\biggl[\left\langle n\left|\nabla_{\mathbf{k}}\boldsymbol{\Theta}_{\mathbf{k}}\right|n\right\rangle + \nonumber \\
 & -i\frac{\lambda}{a}\omega_{n\mathbf{k}}^{2}\left(\left\langle \nabla_{\mathbf{k}}n|\partial_{k_{x}}n\right\rangle -\left\langle \partial_{k_{x}}n|\nabla_{\mathbf{k}}n\right\rangle \right)+\nonumber \\
 & \left.+i\frac{\lambda}{a}\omega_{n\mathbf{k}}^{2}\left(\left\langle \nabla_{\mathbf{k}}n|n\right\rangle \left\langle n| \partial_{k_{x}}n\right\rangle -\left\langle n|\nabla_{\mathbf{k}}n\right\rangle \left\langle \partial_{k_{x}}n|n\right\rangle \right)\right] \, . \label{eq:63} 
\end{alignat}
At last, in close analogy to the perturbative approach described for the electronic transport in the previous section, 
the last line in Eq.~(\ref{eq:63}) gives a null contribution, and with little algebraic effort we obtain
\begin{alignat}{1}
\tilde{\mathit{\mathbf{v}}}_{n\mathbf{k}}^{e}= & \frac{1}{2\omega_{n\mathbf{k}}\left\langle n|n\right\rangle }\left[\left\langle n\left|\nabla_{\mathbf{k}}\boldsymbol{\Theta}_{\mathbf{k}}\right|n\right\rangle -\frac{\lambda}{a}\omega_{n\mathbf{k}}^{2}\hat{\mathbf{y}}\left(\boldsymbol{\Omega}_{n\mathbf{k}}\cdot\hat{\mathbf{z}}\right)\right] \nonumber \\
= & \mathit{\mathbf{v}}_{n\mathbf{k}}^{e}-\frac{\lambda}{2a}\frac{\omega_{n\mathbf{k}}\hat{\mathbf{y}}\left(\boldsymbol{\Omega}_{n\mathbf{k}}\cdot\hat{\mathbf{z}}\right)}{\left\langle n|n\right\rangle } \, .  \label{eq:64}
\end{alignat}
If we now define the generalized wave vector equation
\begin{equation}
\dot{\boldsymbol{\kappa}}\doteq-\frac{\lambda}{2a}\frac{\omega_{n\mathbf{k}}\hat{\mathbf{x}}}{\left\langle n|n\right\rangle } \, ,\label{eq:65}
\end{equation}
we can recast Eq.~(\ref{eq:64}) in the compact and familiar form
\begin{alignat}{1}
\tilde{\mathit{\mathbf{v}}}_{n\mathbf{k}}^{e} & =\mathit{\mathbf{v}}_{n\mathbf{k}}^{e}-\dot{\boldsymbol{\kappa}}\times\boldsymbol{\Omega}_{n\mathbf{k}} \, ,\label{eq:66}
\end{alignat}
which exactly represents the photonic analog of Eq.~(\ref{eq:2}). {In fact, from Eq.~(\ref{eq:63}) a ``photonic'' Berry curvature can be explicitly defined by a formally similar expression to the electronic case, $\boldsymbol{\Omega}_{n\mathbf{k}}\doteq i\left(rot\left\langle n|\nabla_{\mathbf{k}}n\right\rangle \right)$. 
We stress that the derivation of Eq.~(\ref{eq:66}) is strictly valid only within the assumptions made throughout this section, namely considering 2D 
photonic crystals made of non-dispersive gyrotropic materials in which no TE/TM mixing occurs.
We notice that also here the Berry curvature is interpreted as a geometric property of the gauge bundle, 
referred to a given photonic band. 
The relevance of this topological property in photonic systems has been already discussed in the literature \cite{key-5,key-11}. 
In particular, when this quantity is different from zero, the gauge bundle is non-trivial and the topology of the bundle affects the dynamics 
of light propagation in the system. 
As remarked by Raghu and Haldane \cite{key-5}, a necessary condition to obtain a non-trivial bundle is TRS breaking, as in the gyrotropic
2D photonic crystal system assumed here. 
In general, when the degeneracy of photonic bands along high-symmetry direction in reciprocal space is removed by TRS breaking, 
the associated bundle can ``twist'' giving rise to non-trivial topological features, which can manifest themselves, e.g., as ballistic one-way 
edge states \cite{key-11}.

Despite the formal similarities between electronic and photonic periodic systems, there is a considerable number of physical and mathematical 
differences that should be carefully taken into account, such as the vectorial nature of operators and fields or the re-formulation of HF theorem 
in the general case. 
However, it is interesting to emphasize that the detection of such topological features appears easier in the photonic case, since electromagnetic 
edges mode can be observed by direct injection of light into the system, as it is further discussed  in the next section.
 }

\section{Discussion and Physical realization}\label{sec:discussion}

Equation~(\ref{eq:66}) is the central result of this paper. It rigorously
shows that by perturbing a gyrotropic photonic crystal with a linear 
grading of the permittivity (permeability), the expression for the energy 
velocity of an electromagnetic mode contains, in addition to a zero-order
term, a geometrical term depending on the Berry curvature, in full analogy 
with the semiclassical equation for the electron velocity in quantum Hall systems. 
Within this framework, the bulk dynamics of light propagation in a gyrotropic photonic 
crystal is deeply connected to the recent observation of back-scattering immune edge
states \cite{key-12}, which is now rigorously explained in the photonic context by the 
bulk-edge correspondence \cite{key-9} (also known as \textit{holographic principle}).
In fact, although the presence of such edge states in systems with broken TRS
is a clear indication of non-trivial topological properties, a bulk theory is always 
needed to rigorously justify and fully understand their physics.

\begin{figure}[t]
\begin{center}
\vspace{-0.6cm}
\includegraphics[width=0.5\textwidth]{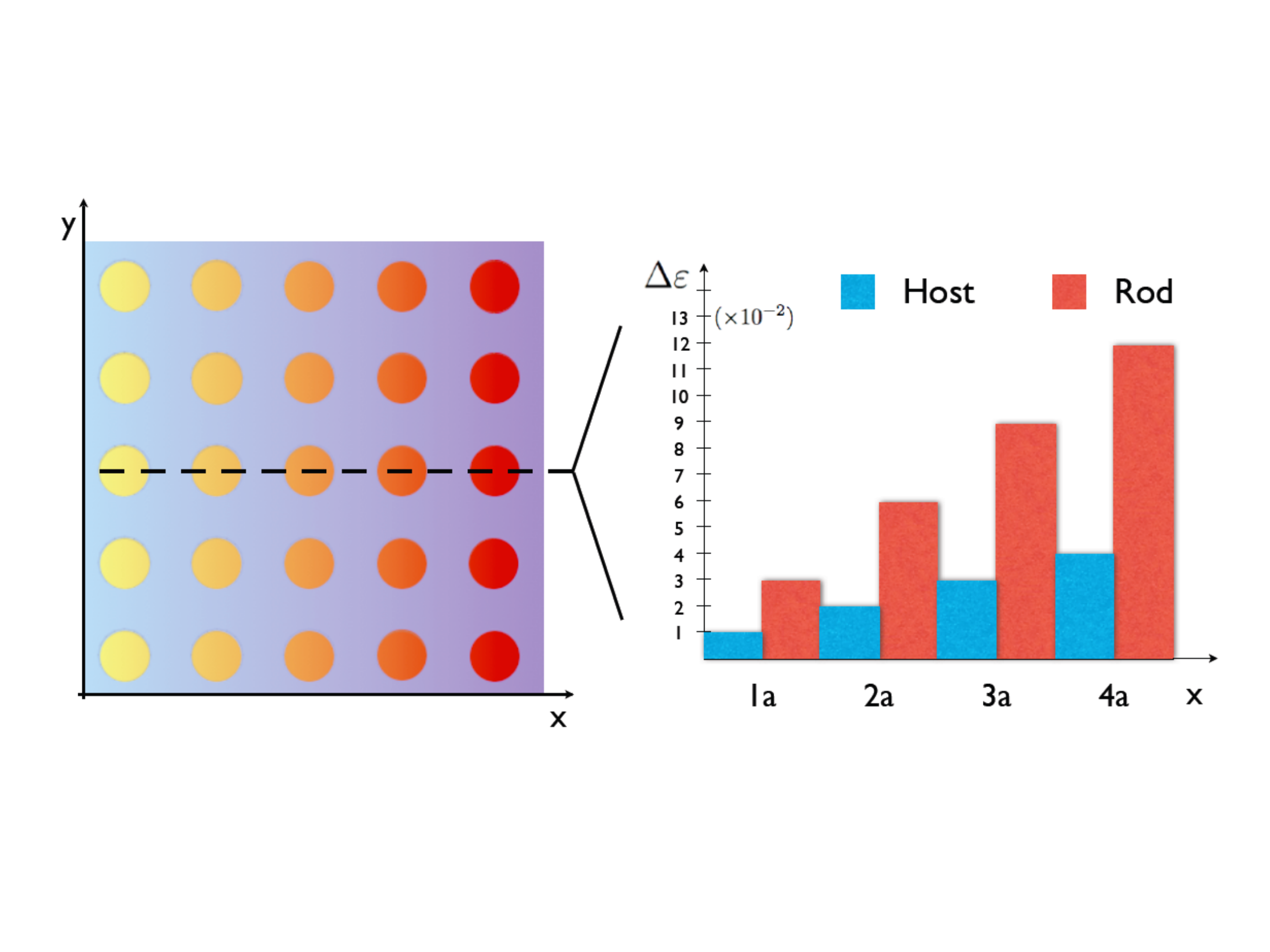}
\vspace{-1.8cm} 
\caption{(Color online)  Schematic of a gyromagnetic pillar-based
photonic crystal with weak index grading along the propagation direction, 
for a bulk photonic crystal analog of the quantum Hall effect 
to be experimentally shown.} \label{fig2}
\end{center}
\end{figure}

A possible experimental scheme to demonstrate the bulk analog of the quantum Hall 
effect is proposed in Fig.~\ref{fig2}. It shows a weak  (e.g., $\lambda\sim 10^{-2}$) 
linear grading of the dielectric permittivity along the propagation direction of a 2D 
YIG-rods photonic crystal. Indeed, grading of the refractive index can be technologically achieved
in different ways today  \cite{cassanOE2012}.
A light beam propagating along the grading direction would experience a topology-related
bending within the photonic crystal region, according to Eq.~(\ref{eq:66}), that would not be 
present in the absence of TRS breaking.

Finally, we would like to stress that a rigorous reformulation
of the semiclassical equation of motion, Eq.~(\ref{eq:3}), expressing 
the time derivative of the crystal momentum $\mathbf{k}$, is out of the range 
of application of the present model. In fact, even if we can speculate that the equation
for $\dot{\mathbf{k}}$ should have the same explicit expression as in 
Eq.~(\ref{eq:65}), the demonstration that $\dot{\mathbf{k}}=\dot{\boldsymbol{\kappa}}$
has to be found independently of the theoretical framework presented so far, 
which goes beyond the scopes of the present work.

% FIG. 2

\section{Conclusions}

In summary, we have derived a perturbative theory of the 
photonic transport in {two-dimensional, non-dispersive photonic crystals with 
gyrotropic constituents and a weak permittivity grading}.
Time-reversal symmetry is broken by the gyrotropic nature of the metamaterials 
employed, in analogy to the magnetic field in the electron transport, while the role 
of the dragging force induced by the electric  field in quantum Hall systems is played 
here by the weak permittivity grading along the propagation direction. 
The specularity of the theoretical formulation between electric and magnetic
fields, in terms of a generalized eigenvalue problem from Maxwell equations, 
allows a direct transfer of these results to two-dimensional photonic crystals 
made of gyroelectric materials with a grading of the magnetic permeability.

Under the assumption made, we have found that a complete formal analogy exists between 
the semiclassical equations of motion for an electron in a quantum Hall system and the 
electromagnetic energy transport in such a bulk photonic crystal, where the energy velocity 
of a given photonic mode is corrected by a topology-related term that has the meaning of a Berry 
curvature. We have rigorously derived the explicit expression of the Berry curvature 
in terms of differential operators derived from Maxwell equations, with a formally analogous
procedure to the electronic case. Thanks to the bulk-edge correspondence, this work 
gives a fully rigorous theoretical account of the recent experimental results obtained for 
electromagnetic energy transport through back-scattering immune chiral edge states.

Moreover, these results allow to design possible experimental configurations where 
the direct photonic analog of the quantum Hall effect can be probed in a bulk two-dimensional 
photonic crystal, instead of edge transport. In fact, the two-dimensional propagation of an 
electromagnetic beam in the photonic crystal region should be strongly influenced
by the topological term, and a measurement of the beam deviation from the expected 
direction should give a direct measure of the Berry curvature in such a topological photonic 
insulator system.

\acknowledgments{
The authors acknowledge useful and stimulating discussions with G. De Nittis. 
DG acknowledges R. Fazio, M. Gibertini, and M. Polini for motivating discussions on 
photonic topological insulators.}

\appendix 

% APPENDIX A
\section{Explicit form of the photonic eigenvalue equation }

Here we explicitly obtain the eigenvalue problem for the field component $F_{z}$ starting from Eq.~(\ref{eq:36}) 
\begin{equation}
\overset{\leftrightarrow}{\boldsymbol{\varepsilon}}{}^{-\frac{1}{2}}\nabla\times\left[\overset{\leftrightarrow}{\boldsymbol{\mu}}^{-1}\nabla\times\left(\overset{\leftrightarrow}{\boldsymbol{\varepsilon}}{}^{-\frac{1}{2}}\boldsymbol{F}_e\right)\right]=\omega^{2}\boldsymbol{F}_e \, .\label{eq:67}
\end{equation}
First, we impose the condition
\[
\boldsymbol{F}=\boldsymbol{F}_e=(0,0,F_{z}) \, ,
\]
which is justified because there is no TE/TM mixing in our system, from the assumption we have made on the tensorial form of the permittivity and permeability tensors, Eqs.~(\ref{eq:40}-\ref{eq:41}), respectively . 
We now calculate Eq.~(\ref{eq:67}) step by step, starting from
\begin{flalign}
\nabla\times\left(\overset{\leftrightarrow}{\boldsymbol{\varepsilon}}{}^{-\frac{1}{2}}\boldsymbol{F}\right) & =\left(-\frac{1}{2}\varepsilon^{-\frac{3}{2}}\partial_{y}\varepsilon F_{z}+\varepsilon^{-\frac{1}{2}}\partial_{y} F_{z}\right)\hat{\mathbf{x}}+\nonumber \\
 & +\left(\frac{1}{2}\varepsilon^{-\frac{3}{2}}\partial_{x}\varepsilon F_{z}-\varepsilon^{-\frac{1}{2}}\partial_{x} F_{z}\right)\hat{\mathbf{y}}\nonumber \\
 & \doteq\zeta_{x}\hat{\mathbf{x}}+\zeta_{y}\hat{\mathbf{y}},
\end{flalign}
with obvious definitions of $\zeta_{x}$ and $\zeta_{y}$. The second step is to calculate
\begin{flalign}
\overset{\leftrightarrow}{\boldsymbol{\mu}}^{-1}\nabla\times\left(\overset{\leftrightarrow}{\boldsymbol{\varepsilon}}{}^{-\frac{1}{2}}\boldsymbol{F}\right)= & \overset{\leftrightarrow}{\boldsymbol{\mu}}^{-1}\boldsymbol{\zeta} \nonumber \\
= & \left(\bar{\mu}^{-1}\zeta_{x}+i\eta\zeta_{y}\right)\hat{\mathbf{x}}+\nonumber \\
 & +\left(\bar{\mu}^{-1}\zeta_{y}-i\eta\zeta_{x}\right)\hat{\mathbf{y}}\nonumber \\
\doteq & \theta_{x}\hat{\mathbf{x}}+\theta_{y}\hat{\mathbf{y}} \, ,  \label{eq:69}  
\end{flalign}
with obvious definitions of $\theta_{x}$ and $\theta_{y}$, as before.
From Eq.~(\ref{eq:69}) we obtain 
\begin{alignat}{1}
\nabla\times\overset{\leftrightarrow}{\boldsymbol{\mu}}^{-1}\nabla\times\left(\overset{\leftrightarrow}{\boldsymbol{\varepsilon}}{}^{-\frac{1}{2}}\boldsymbol{F}\right)= & \nabla\times\boldsymbol{\theta} \nonumber \\
= & \left(-\partial_{z}\theta_{y}\right)\hat{\mathbf{x}}+\left(\partial_{z}\theta_{x}\right)\hat{\mathbf{y}}+\nonumber \\
 & +\left(\partial_{x}\theta_{y}-\partial_{y}\theta_{x}\right)\hat{\mathbf{z}} \, , \label{eq:70} 
\end{alignat}
from which, using the relation $\partial_{z}F_{z}=0$, which derives
from the transversality condition $\nabla\cdot\left(\overset{\leftrightarrow}{\boldsymbol{\varepsilon}}\boldsymbol{{E}}\right)$,
and the relations $\partial_{z}\varepsilon=\partial_{z}\mu=\partial_{z}\eta=0$,
due to the symmetry of the system, we get
\begin{equation}
\nabla\times\overset{\leftrightarrow}{\boldsymbol{\mu}}^{-1}\nabla\times\left(\overset{\leftrightarrow}{\boldsymbol{\varepsilon}}{}^{-\frac{1}{2}}\boldsymbol{F}\right)=\left(\partial_{x}\theta_{y}-\partial_{y}\theta_{x}\right)\hat{\mathbf{z}} \, .\label{eq:71}
\end{equation}
Multiplying Eq.~(\ref{eq:71}) by $\varepsilon^{-\frac{1}{2}}$,
we obtain the eigenvalue problem
\begin{flalign}
\omega^{2}F_{z}= & \left[-\varepsilon^{-1}\bar{\mu}^{-1}\nabla^{2}+\right.\label{eq:72}\\
 & +\varepsilon^{-1}\left(\bar{\mu}^{-1}\varepsilon^{-1}\nabla\varepsilon-\nabla\bar{\mu}^{-1}-i\left(\hat{\mathbf{z}}\times\nabla\eta\right)\right)\cdot\nabla+\nonumber \\
 & +\frac{1}{2}\varepsilon^{-2}\Bigl(\nabla\bar{\mu}^{-1}\cdot\nabla\varepsilon-\frac{3}{2}\bar{\mu}^{-1}\varepsilon^{-1}\left(\nabla\varepsilon\right)^{2}\nonumber \\
 & \left.+i\left(\hat{\mathbf{z}}\times\nabla\eta\right)\cdot\nabla\varepsilon+\bar{\mu}^{-1}\nabla^{2}\varepsilon\Bigr)\right] F_{z} \, .\nonumber 
\end{flalign}
Using the relations 
\begin{eqnarray}
\varepsilon^{-2}\nabla\varepsilon & = & -\nabla\varepsilon^{-1}\label{eq:73}\\
\varepsilon^{-2}\nabla^{2}\varepsilon & = & 2\varepsilon\left(\nabla\varepsilon^{-1}\right)^{2}-\nabla^{2}\varepsilon^{-1},\label{eq:74}
\end{eqnarray}
to point out the role of $\varepsilon^{-1}$ with respect to 
$\varepsilon$,  Eq.~(\ref{eq:72}) assumes exactly the same expression
as in Eq.~(\ref{eq:43}) with the operator in Eq.~(\ref{eq:43bis}).

% APPENDIX B
\section{Explicit derivation of the perturbed photonic operator}

The perturbed operator, obtained by replacing $\varepsilon\rightarrow\varepsilon+\Delta\varepsilon$
in Eq.~(\ref{eq:43bis}), has the form
\begin{flalign}
\tilde{\boldsymbol{\Theta}}= & \biggl[-(\varepsilon+\Delta\varepsilon)^{-1}\bar{\mu}^{-1}\nabla^{2}+ \nonumber  \\
 & -\Bigl(\bar{\mu}^{-1}\nabla\left(\varepsilon+\Delta\varepsilon\right)^{-1}+\left(\varepsilon+\Delta\varepsilon\right)^{-1}\nabla\bar{\mu}^{-1}+\nonumber \\
 & +i\left(\varepsilon+\Delta\varepsilon\right)^{-1}\left(\hat{\mathbf{z}}\times\nabla\eta\right)\Bigr)\cdot\nabla+\nonumber \\
 & -\frac{1}{2}\nabla\bar{\mu}^{-1}\cdot\nabla\left(\varepsilon+\Delta\varepsilon\right)^{-1}-\frac{1}{2}\bar{\mu}^{-1}\nabla^{2}\left(\varepsilon+\Delta\varepsilon\right)^{-1}+\nonumber \\
 & +\frac{1}{4}\bar{\mu}^{-1}\left(\varepsilon+\Delta\varepsilon\right)\left(\nabla\left(\varepsilon+\Delta\varepsilon\right)^{-1}\right)^{2}\nonumber \\
 & \left.-\frac{1}{2}i\left(\hat{\mathbf{z}} \times\nabla\eta\right)\cdot\nabla\left(\varepsilon+\Delta\varepsilon\right)^{-1}\right] \, .\label{eq:75}
\end{flalign}
Using the following Taylor expansions
\begin{flalign}
\left(\varepsilon+\Delta\varepsilon\right)^{-1} & \simeq\varepsilon^{-1}-\varepsilon^{-1}\left(\frac{\Delta\varepsilon}{\varepsilon}\right) \nonumber \\
\nabla\left(\varepsilon+\Delta\varepsilon\right)^{-1} & \simeq\nabla\varepsilon^{-1}-\nabla\varepsilon^{-1}\left(\frac{\Delta\varepsilon}{\varepsilon}\right)\nonumber \\
\left(\nabla\left(\varepsilon+\Delta\varepsilon\right)^{-1}\right)^{2} & \simeq\left(\nabla\varepsilon^{-1}\right)^{2}-2\left(\nabla\varepsilon^{-1}\right)^{2}\left(\frac{\Delta\varepsilon}{\varepsilon}\right)\nonumber \\
\nabla^{2}\left(\varepsilon+\Delta\varepsilon\right)^{-1} & \simeq\nabla\left(\nabla\varepsilon^{-1}-\nabla\varepsilon^{-1}\left(\frac{\Delta\varepsilon}{\varepsilon}\right)\right)\nonumber \\
 & \simeq\nabla^{2}\varepsilon^{-1}-\nabla^{2}\varepsilon^{-1}\left(\frac{\Delta\varepsilon}{\varepsilon}\right) \, , \label{eq:76} 
\end{flalign}
which can be obtained assuming that ${\Delta\varepsilon}/{\varepsilon}\ll1$
and $\nabla({\Delta\varepsilon}/{\varepsilon})\simeq 0$, Eq.~(\ref{eq:75}) becomes
\begin{flalign}
\tilde{\boldsymbol{\Theta}}= & \biggl[-\varepsilon^{-1}\bar{\mu}^{-1}\nabla^{2}+ \nonumber \\
 & -\left(\bar{\mu}^{-1}\nabla\varepsilon^{-1}+\varepsilon^{-1}\nabla\bar{\mu}^{-1}+i\varepsilon^{-1}\left(\hat{\mathbf{z}}\times\nabla\eta\right)\right)\cdot\nabla+\nonumber \\
 & -\frac{1}{2}\nabla\bar{\mu}^{-1}\cdot\nabla\varepsilon^{-1}-\frac{1}{2}\bar{\mu}^{-1}\nabla^{2}\varepsilon^{-1}+\nonumber \\
 & \left.+\frac{1}{4}\bar{\mu}^{-1}\varepsilon\left(\nabla\varepsilon^{-1}\right)^{2}-\frac{1}{2}i\left(\hat{\mathbf{z}}\times\nabla\eta\right)\cdot\nabla\varepsilon^{-1}\right]+\nonumber \\
 & -\left(\frac{\Delta\varepsilon}{\varepsilon}\right)\biggl[-\varepsilon^{-1}\bar{\mu}^{-1}\nabla^{2}+\nonumber \\
 & -\left(\bar{\mu}^{-1}\nabla\varepsilon^{-1}+\varepsilon^{-1}\nabla\bar{\mu}^{-1}+i\varepsilon^{-1}\left(\hat{\mathbf{z}}\times\nabla\eta\right)\right)\cdot\nabla+\nonumber \\
 & -\frac{1}{2}\nabla\bar{\mu}^{-1}\cdot\nabla\varepsilon^{-1}-\frac{1}{2}\bar{\mu}^{-1}\nabla^{2}\varepsilon^{-1}+\nonumber \\
 & \left.+\frac{1}{4}\bar{\mu}^{-1}\varepsilon\left(\nabla\varepsilon^{-1}\right)^{2}-\frac{1}{2}i\left(\hat{\mathbf{z}}\times\nabla\eta\right)\cdot\nabla\varepsilon^{-1}\right],\nonumber \\
 & =\boldsymbol{\Theta}-\left(\frac{\Delta\varepsilon}{\varepsilon}\right)\boldsymbol{\Theta} \, , \label{eq:77} 
\end{flalign}
which demonstrates the formal expression given in Eq.~(\ref{eq:45}).

% APPENDIX C
\section{Formulation of Hellmann-Feynman equations in photonic crystal context}

In this appendix we show how the HF theorem is easily reformulated in the photonic crystal context, and explicitly derive the two HF equations that are the photonic crystal analog of Eqs.~(\ref{eq:16}) and (\ref{eq:17}), respectively. We begin by considering the parametric eigenvalue problem for the periodic part of the Bloch function
\begin{alignat}{1}
\boldsymbol{\Theta}_{\mathbf{k}}\left|\boldsymbol{u}_{n\mathbf{k}}\right\rangle = & \omega_{n\mathbf{k}}^{2}\left|\boldsymbol{u}_{n\mathbf{k}}\right\rangle \, .\label{eq:78}
\end{alignat}
Taking the derivative with respect to $\mathbf{k}$ and multiplying both sides by $\left\langle \boldsymbol{u}_{n\mathbf{k}}\right|$ , we obtain
\begin{alignat}{1}
\bigl\langle \boldsymbol{u}_{n\mathbf{k}}\bigr|\nabla_{\mathbf{k}}\boldsymbol{\Theta}_{\mathbf{k}}\bigl|\boldsymbol{u}_{n\mathbf{k}}\bigr\rangle & =\bigl\langle \boldsymbol{u}_{n\mathbf{k}}\bigr|\nabla_{\mathbf{k}}\omega_{n\mathbf{k}}^{2}\bigl|\boldsymbol{u}_{n\mathbf{k}}\bigr\rangle \nonumber \\
 & =2\omega_{n\mathbf{k}}\nabla_{\mathbf{k}}\omega_{n\mathbf{k}}\bigl\langle \boldsymbol{u}_{n\mathbf{k}}\bigl|\boldsymbol{u}_{n\mathbf{k}}\bigr\rangle \, , \label{eq:79} 
\end{alignat}
from which, given the definition of group velocity as $\mathit{\mathbf{v}}_{n\mathbf{k}}^{(g)}=\nabla_{\mathbf{k}}\omega_{n\mathbf{k}}$,
we get
\begin{equation}
\mathit{\mathbf{v}}_{n\mathbf{k}}^{(g)}=\frac{1}{2\omega_{n\mathbf{k}}}\frac{\bigl\langle \boldsymbol{u}_{n\mathbf{k}}\bigr|\nabla_{\mathbf{k}}\boldsymbol{\Theta}_{\mathbf{k}}\bigl|\boldsymbol{u}_{n\mathbf{k}}\bigr\rangle}{\bigl\langle \boldsymbol{u}_{n\mathbf{k}}\bigl|\boldsymbol{u}_{n\mathbf{k}}\bigr\rangle} \, ,\label{eq:80}
\end{equation}
which is the photonic formulation of the HF theorem reported in Eq.~(\ref{eq:13}).
In the same way, by differentiating Eq.~(\ref{eq:78})  with respect to $\mathbf{k}$, and then 
multiplying by $\left\langle \boldsymbol{u}_{m\mathbf{k}}\right|$, we get
\begin{equation}
\bigl\langle\boldsymbol{u}_{m\mathbf{k}}\bigr|\nabla_{\mathbf{k}}\boldsymbol{\Theta}_{\mathbf{k}}\bigl|\boldsymbol{u}_{n\mathbf{k}}\bigr\rangle=\left(\omega_{n\mathbf{k}}^{2}-\omega_{m\mathbf{k}}^{2}\right)\bigl\langle\boldsymbol{u}_{m\mathbf{k}}\bigr|\nabla_{\mathbf{k}}\boldsymbol{u}_{n\mathbf{k}}\bigr\rangle \, ,\label{eq:81}
\end{equation}
which is the photonic formulation of the first HF equation reported in Eq.~(\ref{eq:16}). 

Unfortunately, the tensorial nature of the operators makes the photonic reformulation of the second 
HF equation, Eq.~(\ref{eq:17}) rather difficult to demonstrate explicitly in the general case. However, we show here a demonstration for the
particular case considered, i.e. applying the photonic operator to the scalar component of the field. 
From Eq.~(\ref{eq:43}) with Eq.~(\ref{eq:43bis}) we have
\begin{widetext}
\begin{alignat}{1}
\boldsymbol{\Theta}_{\mathbf{k}}u_{n\mathbf{k}}= & \biggl[-\varepsilon^{-1}\bar{\mu}^{-1}\nabla^{2} -\left(2i\varepsilon^{-1}\bar{\mu}^{-1}\mathbf{k}+\bar{\mu}^{-1}\nabla\varepsilon^{-1}+\varepsilon^{-1}\nabla\bar{\mu}^{-1}+i\varepsilon^{-1}\left(\hat{\mathbf{z}}\times\nabla\eta\right)\right)\cdot\nabla+\nonumber \\
 & +\varepsilon^{-1}\bar{\mu}^{-1}k^{2}-i\bar{\mu}^{-1}\mathbf{k}\cdot\nabla\varepsilon^{-1}-i\varepsilon^{-1}\mathbf{k}\cdot\nabla\bar{\mu}^{-1}+\varepsilon^{-1}\mathbf{k}\cdot\left(\hat{\mathbf{z}}\times\nabla\eta\right)+\nonumber \\
 & -\frac{1}{2}i\left(\hat{\mathbf{z}}\times\nabla\eta\right)\cdot\nabla\varepsilon^{-1}-\frac{1}{2}\nabla\bar{\mu}^{-1}\cdot\nabla\varepsilon^{-1}+\frac{1}{4}\bar{\mu}^{-1}\varepsilon\left(\nabla\varepsilon^{-1}\right)^{2}+\nonumber \\
 & \left.-\frac{1}{2}\bar{\mu}^{-1}\nabla^{2}\varepsilon^{-1}\right]u_{n\mathbf{k}}=\omega_{n\mathbf{k}}^{2}u_{n\mathbf{k}} \, , \label{eq:82} 
\end{alignat}
\end{widetext}
where $u_{n\mathbf{k}}$ is the Bloch part of the component $F_z$, and hence 
\begin{flalign}
\nabla_{\mathbf{k}}\boldsymbol{\Theta}_{\mathbf{k}}= & \nabla_{\mathbf{k}}\left[\varepsilon^{-1}\bar{\mu}^{-1}(-i\nabla+\mathbf{k})^{2}-i\bar{\mu}^{-1}\nabla\varepsilon^{-1}+\right.\nonumber \\
 & \left.-i\varepsilon^{-1}\nabla\bar{\mu}^{-1}+\varepsilon^{-1}\left(\hat{\mathbf{z}}\times\nabla\eta\right)\right]\nonumber \\
= & -2i\varepsilon^{-1}\bar{\mu}^{-1}\nabla+2\varepsilon^{-1}\bar{\mu}^{-1}\mathbf{k}-i\left(\bar{\mu}^{-1}\nabla\varepsilon^{-1}+\right.\nonumber \\
 & \left.+\varepsilon^{-1}\nabla\bar{\mu}^{-1}+i\varepsilon^{-1}\left(\hat{\mathbf{z}}\times\nabla\eta\right)\right) \, . \label{eq:83} 
\end{flalign}
Using now the commutation relations
\begin{eqnarray}
\left[\nabla,\mathbf{r}\right] & = & 1\label{eq:84}\\
\left[\nabla^{2},\mathbf{r}\right] & = & 2\nabla,\label{eq:85}
\end{eqnarray}
it is straightforward to show 
\begin{flalign}
\left[\boldsymbol{\Theta}_{\mathbf{k}},\mathbf{r}\right]= & -2\varepsilon^{-1}\bar{\mu}^{-1}\nabla-2i\varepsilon^{-1}\bar{\mu}^{-1}\mathbf{k}-\bar{\mu}^{-1}\nabla\varepsilon^{-1}+\nonumber \\
 & -\varepsilon^{-1}\nabla\bar{\mu}^{-1}-i\varepsilon^{-1}\left(\hat{\mathbf{z}}\times\nabla\eta\right)=-i\nabla_{\mathbf{k}}\boldsymbol{\Theta}_{\mathbf{k}} \, . \label{eq:86} 
\end{flalign}
Using the last relation in Eq.~(\ref{eq:81}), we obtain
\begin{alignat}{1}
i\left\langle u_{m\mathbf{k}}\right|\left[\boldsymbol{\Theta}_{\mathbf{k}},\mathbf{r}\right]\left|u_{n\mathbf{k}}\right\rangle  & =\left(\omega_{n\mathbf{k}}^{2}-\omega_{m\mathbf{k}}^{2}\right)\left\langle u_{m\mathbf{k}}|\nabla_{\mathbf{k}}u_{n\mathbf{k}}\right\rangle \,  \label{eq:87}
\end{alignat}
from which
\begin{alignat}{1}
i\left\langle u_{m\mathbf{k}}\right|\boldsymbol{\Theta}_{\mathbf{k}}\mathbf{r}-\mathbf{r}\boldsymbol{\Theta}_{\mathbf{k}}\left|u_{n\mathbf{k}}\right\rangle  & =i\left\langle u_{m\mathbf{k}}\right|\mathbf{r}\left|u_{n\mathbf{k}}\right\rangle \left(\omega_{m\mathbf{k}}^{2}-\omega_{n\mathbf{k}}^{2}\right)\nonumber \\
 & =\left(\omega_{n\mathbf{k}}^{2}-\omega_{m\mathbf{k}}^{2}\right)\left\langle u_{m\mathbf{k}}|\nabla_{\mathbf{k}}u_{n\mathbf{k}}\right\rangle \, , \label{eq:88} 
\end{alignat}
which finally gives
\begin{equation}
\left\langle u_{m\mathbf{k}}\right|\mathbf{r}\left|u_{n\mathbf{k}}\right\rangle =i\left\langle u_{m\mathbf{k}}|\nabla_{\mathbf{k}}u_{n\mathbf{k}}\right\rangle \, ,\label{eq:89}
\end{equation}
i.e. exactly the photonic crystal analog of Eq.~(\ref{eq:17}).

% APPENDIX D
\section{Demonstration of an expression for the energy velocity}

In the perturbed two-dimensional photonic crystal considered, {where separation of TE/TM modes occurs and the permittivity/permeability tensors are assumed non-dispersive}, from Eq.~(\ref{eq:51}) we can write Eq.~(\ref{eq:53}) as 
\begin{equation}
\tilde{\mathit{\mathbf{v}}}_{n\mathbf{k}}^{(e)}=\frac{\mathrm{Re}\int \mathrm{d}^{3}\mathbf{r} \, \, \mathrm{\tilde{E}}_{n\mathbf{k}}^{*(z)}\times\mathrm{\tilde{H}}_{n\mathbf{k}}^{(x,y)}}{\int \mathrm{d}^{3}\mathbf{r} \, \, \varepsilon\bigl|\mathrm{{E}}_{n\mathbf{k}}^{(z)}\bigr|^{2}} \, ,\label{eq:90}
\end{equation}
where $\mathrm{\tilde{E}}_{n\mathbf{k}}^{*(z)}$ and $\mathrm{\tilde{H}}_{n\mathbf{k}}^{(x,y)}$
are the perturbed electric and magnetic fields, respectively.  
{In Eq.~(\ref{eq:90}) we have implicitly kept only the unperturbed product in the denominator, and we considered twice the electric contribution to the total electromagnetic energy density, i.e. $\mathcal{U}_{n\mathbf{k}} = 2 \mathcal{U}_{n\mathbf{k}}^{{e}} $ as it is true for harmonic modes (see, e.g., page 16 of Ref. \onlinecite{key-14}. }
Using the Maxwell equation 
\begin{equation}
\nabla\times\mathbf{E}\left(\mathbf{r}\right)=i\omega\overset{\leftrightarrow}{\boldsymbol{\mu}}(\mathbf{r})\mathbf{H}\left(\mathbf{r}\right) \, ,\label{eq:91}
\end{equation}
we can rewrite Eq.~(\ref{eq:90}) as
\begin{alignat}{1}
\tilde{\mathit{\mathbf{v}}}_{n\mathbf{k}}^{(e)}= & \frac{1}{\omega\int \mathrm{d}^{3}\mathbf{r} \, \, \varepsilon\bigl|\mathrm{{E}}_{n\mathbf{k}}^{(z)}\bigr|^{2}}\cdot  \nonumber \\
 & \mathrm{Re}\int \mathrm{d}^{3}\mathbf{r} \, \left[-i\bar{\mu}^{-1}\mathrm{\tilde{E}}_{n\mathbf{k}}^{*(z)}\nabla\mathrm{\tilde{E}}_{n\mathbf{k}}^{(z)}-\eta\mathrm{\tilde{E}}_{n\mathbf{k}}^{*(z)}\left(\hat{\mathbf{z}}\times\nabla\mathrm{\tilde{E}}_{n\mathbf{k}}^{(z)}\right)\right]  \, . \label{eq:92} 
\end{alignat}
From Eq.~(\ref{eq:48}) we straightforwardly obtain the relation 
\begin{alignat}{1}
\bigl|\mathrm{\tilde{E}}_{n\mathbf{k}}^{(z)}\bigr\rangle & =\bigl|\mathrm{E}_{n\mathbf{k}}^{(z)}\bigr\rangle+\sum_{m\neq n}\bigl|\mathrm{E}_{m\mathbf{k}}^{(z)}\bigr\rangle\frac{\left\langle \varepsilon^{\frac{1}{2}}\mathrm{E}_{m\mathbf{k}}^{(z)}|\boldsymbol{V}^{p}\left(\varepsilon^{\frac{1}{2}}\mathrm{E}_{n\mathbf{k}}^{(z)}\right)\right\rangle }{\omega_{n\mathbf{k}}^{2}-\omega_{m\mathbf{k}}^{2}}\nonumber \\
 & =\bigl|\mathrm{E}_{n\mathbf{k}}^{(z)}\bigr\rangle+\sum_{m\neq n}\mathrm{J}_{mn}\bigl|\mathrm{E}_{m\mathbf{k}}^{(z)}\bigr\rangle \, , \label{eq:93} 
\end{alignat}
which, once inserted in Eq.~(\ref{eq:92}), gives 
\begin{alignat}{1}
\tilde{\mathit{\mathbf{v}}}_{n\mathbf{k}}^{(e)}= & \mathit{\mathbf{v}}_{n\mathbf{k}}^{(e)}+\frac{1}{\omega\int \mathrm{d}^{3}\mathbf{r} \, \varepsilon\bigl|\mathrm{{E}}_{n\mathbf{k}}^{(z)}\bigr|^{2}}\cdot \nonumber \\
 & \cdot\mathrm{Re}\int \mathrm{d}^{3}\mathbf{r} \, \sum_{m\neq n}\left[-i\bar{\mu}^{-1}\mathrm{J}_{mn}^{*}\mathrm{E}_{m\mathbf{k}}^{*(z)}\nabla\mathrm{E}_{n\mathbf{k}}^{(z)}+\right.\nonumber \\
 & -i\bar{\mu}^{-1}\mathrm{J}_{mn}\mathrm{E}_{n\mathbf{k}}^{*(z)}\nabla\mathrm{E}_{m\mathbf{k}}^{(z)}-\eta\mathrm{J}_{mn}^{*}\mathrm{E}_{m\mathbf{k}}^{*(z)}\left(\hat{\mathbf{z}}\times\nabla\mathrm{E}_{n\mathbf{k}}^{(z)}\right)+\nonumber \\
 & \left.-\eta\mathrm{J}_{mn}\mathrm{E}_{n\mathbf{k}}^{*(z)}\left(\hat{\mathbf{z}}\times\nabla\mathrm{E}_{m\mathbf{k}}^{(z)}\right)\right] \, . \label{eq:94}  
\end{alignat}
Integrating the second and the forth term in the square brackets by parts,
the last equation becomes
\begin{alignat}{1}
\tilde{\mathit{\mathbf{v}}}_{n\mathbf{k}}^{(e)}= & \mathit{\mathbf{v}}_{n\mathbf{k}}^{(e)}+\frac{1}{\omega\int \mathrm{d}^{3}\mathbf{r} \, \varepsilon\bigl|\mathrm{{E}}_{n\mathbf{k}}^{(z)}\bigr|^{2}}\cdot\nonumber \\
 & \cdot\mathrm{Re}\int \mathrm{d}^{3}\mathbf{r}\sum_{m\neq n}\left[-2\bar{\mu}^{-1}\mathrm{Re}\left(i\mathrm{J}_{mn}^{*}\mathrm{E}_{m\mathbf{k}}^{*(z)}\nabla\mathrm{E}_{n\mathbf{k}}^{(z)}\right)+\right.\nonumber \\
 & +i\nabla\bar{\mu}^{-1}\mathrm{J}_{mn}\mathrm{E}_{n\mathbf{k}}^{*(z)}\mathrm{E}_{m\mathbf{k}}^{(z)}+\nonumber \\
 & -2i\eta\mathrm{Im}\left(\mathrm{J}_{mn}^{*}\mathrm{E}_{m\mathbf{k}}^{*(z)}\left(\hat{\mathbf{z}}\times\nabla\mathrm{E}_{n\mathbf{k}}^{(z)}\right)\right)+\nonumber \\
 & \left.+\mathrm{J}_{mn}\mathrm{E}_{n\mathbf{k}}^{*(z)}\mathrm{E}_{m\mathbf{k}}^{(z)}\left(\hat{\mathbf{z}}\times\nabla\eta\right)\right] \, . \label{eq:95} 
\end{alignat}
Except for the first term, all the other terms in the square bracket of
Eq.~(\ref{eq:95}) give a null contribution. In fact, after observing
that $\mathrm{J}_{mn}$ does not depend on spatial variables, we can
see that the third term is purely imaginary, while the second and the
fourth are odd functions \cite{foot4}. Thus, we can write
\begin{alignat}{1}
\tilde{\mathit{\mathbf{v}}}_{n\mathbf{k}}^{(e)}= & \mathit{\mathbf{v}}_{n\mathbf{k}}^{(e)}-\frac{2}{\omega\int \mathrm{d}^{3}\mathbf{r} \, \varepsilon\bigl|\mathrm{E}_{n\mathbf{k}}^{(z)}\bigr|^{2}}\cdot \nonumber\\
 & \cdot\mathrm{Re}\int \mathrm{d}^{3}\mathbf{r}\sum_{m\neq n}\bar{\mu}^{-1}\mathrm{Re}\left(i\mathrm{J}_{mn}^{*}\mathrm{E}_{m\mathbf{k}}^{*(z)}\nabla\mathrm{E}_{n\mathbf{k}}^{(z)}\right) \, . \label{eq:96} 
\end{alignat}

Now we have to verify if the following equality is correct
\begin{alignat}{1}
\tilde{\mathit{\mathbf{v}}}_{n\mathbf{k}}^{(e)}\overset{?}{=} & \frac{1}{2\omega}\frac{\left\langle \tilde{u}_{n\mathbf{k}} \left|\nabla_{\mathbf{k}}\boldsymbol{\Theta}_{\mathbf{k}}\right|\tilde{u}_{n\mathbf{k}}\right\rangle }{\left\langle {u}_{n\mathbf{k}}| {u}_{n\mathbf{k}}\right\rangle } \, ,\label{eq:97}
\end{alignat}
where, as before, $\tilde{u}_{n\mathbf{k}}$ is the Bloch part of the perturbed field component $\tilde{F}_{n\mathbf{k}}=\tilde{F}_{e}^{(z)}$.
To this end, the second member of Eq.~(\ref{eq:97}), which we define $\mathsf{S}$,  can be re-written using the relations
\begin{alignat}{1}
\mathsf{S} & =\frac{1}{2\omega}\frac{\left\langle \tilde{F}_{n\mathbf{k}}\left|\nabla_{\mathbf{k}}\boldsymbol{\Theta}\right|\tilde{F}_{n\mathbf{k}}\right\rangle }{\left\langle {F}_{n\mathbf{k}} | {F}_{n\mathbf{k}}\right\rangle } \nonumber \\
 & =\frac{1}{2\omega}\frac{\left\langle \varepsilon^{\frac{1}{2}}\mathrm{\tilde{E}}_{n\mathbf{k}}^{(z)}\left|\nabla_{\mathbf{k}}\boldsymbol{\Theta}\right|\varepsilon^{\frac{1}{2}}\mathrm{\tilde{E}}_{n\mathbf{k}}^{(z)}\right\rangle }{\left\langle \varepsilon^{\frac{1}{2}}\mathrm{{E}}_{n\mathbf{k}}^{(z)}|\varepsilon^{\frac{1}{2}}\mathrm{{E}}_{n\mathbf{k}}^{(z)}\right\rangle } \, , \label{eq:98} 
\end{alignat}
where we have defined the operator $\nabla_{\mathbf{k}}\boldsymbol{\Theta}$ as
\begin{alignat}{1}
\nabla_{\mathbf{k}}\boldsymbol{\Theta}\bigl|\boldsymbol{F}_{n \mathbf{k}}\bigr\rangle\doteq\nabla_{\mathbf{k}}\boldsymbol{\Theta}_{\mathbf{k}}\bigl|\boldsymbol{u}_{n\mathbf{k}}\bigr\rangle \, ,\label{eq:99}
\end{alignat}
which explicitly gives
\begin{alignat}{1}
\nabla_{\mathbf{k}}\boldsymbol{\Theta}= & -2i\varepsilon^{-1}\bar{\mu}^{-1}\nabla-i\left(\bar{\mu}^{-1}\nabla\varepsilon^{-1}+\right.  \nonumber \\
 & \left.+\varepsilon^{-1}\nabla\bar{\mu}^{-1}+i\varepsilon^{-1}\left(\hat{\mathbf{z}}\times\nabla\eta\right)\right) \, . \label{eq:100} 
\end{alignat}
Using Eqs. (\ref{eq:98}) and (\ref{eq:100}), we obtain
\begin{alignat}{1}
\mathsf{S}= & \frac{1}{2\omega\left\langle \mathrm{{E}}_{n\mathbf{k}}^{(z)}|\varepsilon\mathrm{{E}}_{n\mathbf{k}}^{(z)}\right\rangle }\cdot \nonumber \\
 & \cdot\int \mathrm{d}^{3}\mathbf{r}\left[\mathrm{\tilde{E}}_{n\mathbf{k}}^{*(z)}\left(-i\varepsilon^{-1}\bar{\mu}^{-1}\nabla\varepsilon\mathrm{\tilde{E}}_{n\mathbf{k}}^{(z)}-2i\bar{\mu}^{-1}\nabla\mathrm{\tilde{E}}_{n\mathbf{k}}^{(z)}+\right.\right.\nonumber \\
 & \left.\left.-i\bar{\mu}^{-1}\varepsilon\nabla\varepsilon^{-1}\mathrm{\tilde{E}}_{n\mathbf{k}}^{(z)}-i\nabla\bar{\mu}^{-1}\mathrm{\tilde{E}}_{n\mathbf{k}}^{(z)}+\left(\hat{\mathbf{z}}\times\nabla\eta\right)\mathrm{\tilde{E}}_{n\mathbf{k}}^{(z)}\right)\right] \, , \label{eq:101} 
\end{alignat}
from which, observing that $\nabla\varepsilon^{-1}=-\varepsilon^{-2}\nabla\varepsilon$,
we get
\begin{alignat}{1}
\mathsf{S}= & \frac{1}{2\omega\left\langle {\mathrm{E}}_{n\mathbf{k}}^{(z)}|\varepsilon {\mathrm{E}}_{n\mathbf{k}}^{(z)}\right\rangle }\int \mathrm{d}^{3}\mathbf{r}\left[-2i\bar{\mu}^{-1}\tilde{\mathrm{E}}_{n\mathbf{k}}^{*(z)}\nabla\tilde{\mathrm{E}}_{n\mathbf{k}}^{(z)}+\right. \nonumber \\
 & \left.-i\nabla\bar{\mu}^{-1}\left|\tilde{\mathrm{E}}_{n\mathbf{k}}^{(z)}\right|^{2}+\left(\hat{\mathbf{z}}\times\nabla\eta\right)\left|\tilde{\mathrm{E}}_{n\mathbf{k}}^{(z)}\right|^{2}\right] \, .\label{eq:102} 
\end{alignat}
In the last equation, we can use  the same considerations as before regarding odd functions, from which we get
\begin{alignat}{1}
\mathsf{S}= & \mathbf{v}_{n}^{(e)}(\mathbf{k})-\frac{4}{2\omega\left\langle{\mathrm{E}}_{n\mathbf{k}}^{(z)}|\varepsilon {\mathrm{E}}_{n\mathbf{k}}^{(z)}\right\rangle }\cdot \nonumber \\
 & \cdot\mathrm{Re}\int \mathrm{d}^{3}\mathbf{r}\sum_{m\neq n}\bar{\mu}^{-1}\mathrm{Re}\left(i\mathrm{J}_{mn}^{*}\mathrm{E}_{m\mathbf{k}}^{*(z)}\nabla\mathrm{E}_{n\mathbf{k}}^{(z)}\right) \, ,\label{eq:103} 
\end{alignat}
which compared to Eq. (\ref{eq:96}) finally gives Eq.~(\ref{eq:53}).
%
%
%  REFERENCES
%


\begin{thebibliography}{99}
\bibitem{foot1}
Here we specifically refer to two-dimensional samples with one-dimensional edges, 
as typical of quantum Hall systems.
\bibitem{foot3}
{We are assuming hermitian and positive-defined matrices, hence the square roots of the permittivity and permeability 
tensors exist and are mathematically well defined \cite{key-27,key-28}.}
\bibitem{foot4}
Here, $\nabla\bar{\mu}^{-1}$ and $\nabla\eta$ are periodic odd functions, while terms like $\mathrm{E}_{n\mathbf{k}}^{*(z)}\mathrm{E}_{m\mathbf{k}}^{(z)}$
are periodic because of the Bloch-Floquet theorem.
\end{thebibliography}
\end{document}